\documentclass[lettersize,journal]{IEEEtran}

\usepackage{amsmath,amsfonts}
\usepackage{algorithmic}
\usepackage{algorithm}
\usepackage{array}
\usepackage[caption=false,font=normalsize,labelfont=sf,textfont=sf]{subfig}
\usepackage{textcomp}
\usepackage{url}
\usepackage{verbatim}
\usepackage{graphicx}
\usepackage{cite}

\usepackage{newfloat}
\usepackage{listings}
\usepackage{makecell}
\usepackage{multirow}
\usepackage{rotating}

\usepackage{amsthm}
\theoremstyle{definition}
\newtheorem{definition}{Definition}

\usepackage{booktabs}
\usepackage{arydshln}

\usepackage{colortbl}
\usepackage{microtype}
\usepackage{ctable}
\usepackage{threeparttable}
\usepackage{dblfloatfix}
\usepackage{diagbox}
\usepackage{ragged2e}
\usepackage{url}
\usepackage{lipsum}
\usepackage{xcolor}
\usepackage{listings}

\usepackage{epstopdf}
\usepackage{ragged2e}

\usepackage{iitem}

\begin{document}

\title{Behavior-aware Account De-anonymization\\on Ethereum Interaction Graph}

\author{Jiajun Zhou,
        Chenkai Hu,
		Jianlei Chi,
		Jiajing Wu,~\IEEEmembership{Senior Member,~IEEE},\\
		Meng Shen,~\IEEEmembership{Member,~IEEE},
		and~Qi~Xuan,~\IEEEmembership{Senior Member,~IEEE}
\thanks{J. Zhou and C. Hu are with the Institute of Cyberspace Security, College of Information Engineering, Zhejiang University of Technology, Hangzhou 310023, China. E-mail: jjzhou@zjut.edu.cn, ckhu0122@gmail.com.}
\thanks{J. Chi is with the Hangzhou Research Institute of Xidian University, Hangzhou 311231, China. E-mail: chijianlei@gmail.com.}
\thanks{J. Wu is with the School of Computer Science and Engineering, Sun Yat-sen University, Guangzhou 510006, China. E-mail: wujiajing@mail.sysu.edu.cn.}
\thanks{M. Shen is with the School of Cyberspace Science and Technology, Beijing Institute of Technology, Beijing 100081, China, and also
with Peng Cheng Laboratory (PCL), Shenzhen 518066, China. E-mail: shenmeng@bit.edu.cn.}
\thanks{Q. Xuan is with the Institute of Cyberspace Security, College of Information Engineering, Zhejiang University of Technology, Hangzhou 310023, China, with the PCL Research Center of Networks and Communications, Peng Cheng Laboratory, Shenzhen 518000, China, and also with the Utron Technology Co., Ltd. (as Hangzhou Qianjiang Distinguished Expert), Hangzhou 310056, China. E-mail: xuanqi@zjut.edu.cn.}
\thanks{Corresponding author: Qi Xuan.}
}

\markboth{IEEE Transactions on Information Forensics and Security}
{Shell \MakeLowercase{\textit{et al.}}: A Sample Article Using IEEEtran.cls for IEEE Journals}


\maketitle

\begin{abstract}
	Blockchain technology has the characteristics of decentralization, traceability and tamper-proof, which creates a reliable decentralized trust mechanism, further accelerating the development of blockchain finance.
	However, the anonymization of blockchain hinders market regulation, resulting in increasing illegal activities such as money laundering, gambling and phishing fraud on blockchain financial platforms.
	Thus, financial security has become a top priority in the blockchain ecosystem, calling for effective market regulation.
	In this paper, we consider identifying Ethereum accounts from a graph classification perspective, and propose an end-to-end graph neural network framework named \emph{Ethident}, to characterize the behavior patterns of accounts and further achieve account de-anonymization.
	Specifically, we first construct an Account Interaction Graph (AIG) using raw Ethereum data.
	Then we design a hierarchical graph attention encoder named \emph{HGATE} as the backbone of our framework, which can effectively characterize the node-level account features and subgraph-level behavior patterns.
	For alleviating account label scarcity, we further introduce contrastive self-supervision mechanism as regularization to jointly train our framework.
	Comprehensive experiments on Ethereum datasets demonstrate that our framework achieves superior performance in account identification, yielding 1.13\% $\sim$ 4.93\% relative improvement over previous state-of-the-art.
	Furthermore, detailed analyses illustrate the effectiveness of \emph{Ethident} in identifying and understanding the behavior of known participants in Ethereum (e.g. exchanges, miners, etc.), as well as that of the lawbreakers (e.g. phishing scammers, hackers, etc.), which may aid in risk assessment and market regulation.
\end{abstract}


\section{Introduction}
\IEEEPARstart{T}{he} past few years have witnessed the application of blockchain technology in new technological and industrial revolutions, such as cryptocurrency~\cite{miraz2018applications}, financial services~\cite{fanning2016blockchain}, supply chain management~\cite{blossey2019blockchain}, healthcare~\cite{mcghin2019blockchain}, etc.
As a distributed data storage technology, blockchain is decentralized, traceable and tamper-proof, which guarantees the fidelity and security of data recording and generates trust without a third-party notarization.
Benefiting from these characteristics, blockchain has attracted considerable attention and is best known for its crucial role in the field of digital cryptocurrencies, such as Bitcoin and Ethereum.
According to statistics from market analysis sites such as CoinMarketCap\footnote{\url{https://coinmarketcap.com/}}, as of August 2021, about 11,000 types of cryptocurrencies existed, with a total market value of up to 1.9 trillion dollars.

However, blockchain has become a tempting target for hackers and other cybercriminals due to its huge economic value and anonymization.
Each individual has a virtual identity on blockchain unrelated to the real one, called pseudonym.
For instance, in the Ethereum system, the last 20 bytes of the public key hash are used as the account address (i.e., pseudonym).
However, while pseudonymous accounts protect users' privacy, it also provides shelter for illegal transactions, making it difficult for regulators to identify the culprit.
At present, the weak regulation of blockchain platforms has led to endless financial crimes such as money laundering, gambling and phishing scams.
In 2018, a statistical report published by Kaspersky Lab showed that Ether is the most popular digital asset for criminals, and the loss caused by illegal activities on decentralized applications (DApps) has reached 900 million dollars.
Therefore, financial security has become a top priority in the blockchain ecosystem, and it is of great significance to study security strategies for public blockchain in application scenarios such as risk assessment and market regulation.

\begin{figure}[htp]
	\centering
		\includegraphics[width=\linewidth]{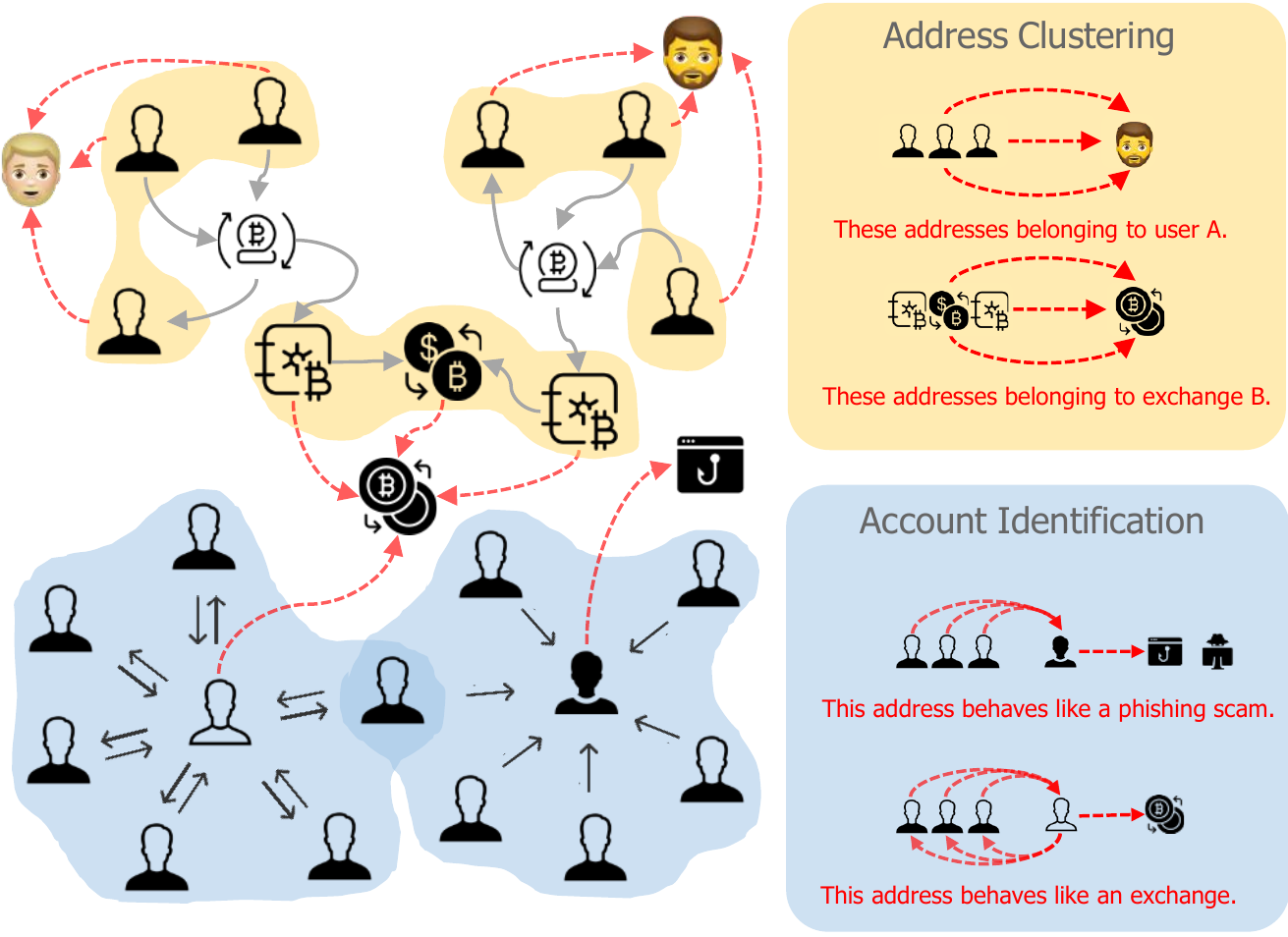}
		\caption{Illustration of the difference between account identification and address clustering.}
		\label{fig: vs}
\end{figure}

\subsection{Account Identification vs. Address Clustering}
Fortunately, the openness and transparency of blockchain make access to block information without barriers.
Recently, existing related work has focused on using the public transaction information to analyze the behavior patterns of accounts and mine the identity information behind them, such as exchanges, phishing scammers, miners and Ponzi schemes, deriving several typical de-anonymization tasks, especially for address clustering and account identification.
Fig.~\ref{fig: vs} shows an illustrative example to explain the difference between address clustering and account identification.
From the definition perspective, address clustering aims to partition the address set observed in Bitcoin transactions into maximal address subsets likely controlled by the same entity, i.e., re-identifying multiple addresses belonging to the same entity.
Account identification aims to determine the identity type of the account by mining the attributes and behavioral characteristics, i.e., attributing the accounts to specific types.
From the task paradigm, the former can be regarded as an unsupervised clustering task, while the latter is generally the supervised classification task.
From the application perspective, a large number of existing address clustering methods are usually designed according to the characteristics of the Bitcoin system, and are usually applied to Bitcoin rather than Ethereum due to their technical differences~\cite{klusman2018deanonymisation}.
While account identification methods only rely on general information such as transaction records on the blockchain, as well as external technologies such as machine learning and network science, thus showing better universality.

\subsection{Challenges}
In this paper, we focus on de-anonymizing Ethereum accounts through account identification.
Existing account identification methods mainly concentrate on manual feature engineering~\cite{toyoda2018multi,lin2019evaluation,bartoletti2018data,huang2020understanding} and graph analytics~\cite{li2020identifying,yuan2020detecting,shen2021identity,chen2020phishing}, which are effective but suffer from several shortcomings and challenges.
First, manual feature engineering relies on the prior knowledge of feature designers and is incapable of capturing the underlying information in blockchain data, such as transaction patterns, resulting in low feature utilization and unsound expressiveness.
In addition, manual features have weak reusability across different blockchain platforms due to technical differences. 
For example, Ethereum data has features associated with smart contracts that Bitcoin does not, which greatly limits the reusability of manual features.
Second, graph analytics relies on large-scale transaction graphs constructed from mass blockchain data, resulting in high computational consumption and time cost when applying graph random walks or graph neural networks (GNNs).
Meanwhile, the growing number of transactions on the blockchain drives frequent updates in the transaction graphs in terms of nodes and edges, which is not conducive to full-graph learning.
Lastly, the annotated information of account identities published in the third-party sites is relatively scarce, resulting in a poor generalization of supervised models. 

\subsection{Our Contributions}
To tackle these challenges, we design a behavior-aware \textbf{Eth}ereum account \textbf{ident}ification framework (\emph{Ethident}) --- an end-to-end graph neural network model, to characterize the behavior patterns of accounts and further achieve account de-anonymization on Ethereum.
Specifically, we first collect and collate large amounts of data involving transaction, smart contract and public annotation of account identity from the Ethereum-related platforms, and then construct an Account Interaction Graph (AIG) and its lightweight version.
Since the large-scale account interaction graph is not feasible for full-batch training of GNNs, we consider account identification as a subgraph-level classification task, and extract neighborhood subgraphs of target accounts from the complete interaction graph, yielding micro interaction subgraphs, which allows for mini-batch training of GNNs.
To better capture the account behavior patterns, we design a \textbf{H}ierarchical \textbf{G}raph \textbf{AT}tention \textbf{E}ncoder named \emph{HGATE} as the backbone of our framework, which can effectively characterize the node-level account features and subgraph-level behavior patterns.
Furthermore, we introduce data augmentation and contrastive self-supervision mechanism for account identification to alleviate the label scarcity that may lead to poor model generalization during supervised learning.
In this way, our framework jointly trains the subgraph contrast and classification tasks, achieving state-of-the-art performance in account identification.
The main contributions of this work are summarized as follows:
\begin{itemize}
	\item \textbf{Data collection:} We construct the Account Interaction Graph (AIG) using collected Ethereum data, and further publish the subgraph datasets for account identification research on Ethereum.
	\item \textbf{Scalability:} We consider identifying Ethereum accounts from a graph classification perspective, and design subgraph sampling strategies to achieve scalable account identification.
	\item \textbf{Powerful feature characterization:} We propose a hierarchical graph attention encoder named \emph{HGATE} to effectively characterize the node-level account features and subgraph-level behavior patterns.
	\item \textbf{Generalization:} We establish a behavior-aware Ethereum account identification framework named \emph{Ethident}\footnote{Data and code are available at https://github.com/jjzhou012/Ethident} which integrates graph augmentation and self-supervision mechanisms, to alleviate the label scarcity and learn highly-expressive behavior pattern representations.
	\item \textbf{State-of-the-art performance:} Extensive experiments on Ethereum datasets demonstrate that our framework can achieve state-of-the-art performance in account identification.
	We further analyze the behavior patterns of different accounts and illustrate the superiority of our framework in terms of performance, scalability and generalization.
\end{itemize}

\begin{table}[htp]
	\centering
	\renewcommand\arraystretch{1.2}
	\caption{Main symbols used in this paper.}
	\label{tb: symbol}
	\resizebox{\linewidth}{!}{%
	\begin{tabular}{lr} 
		\hline\hline
		Symbol                          & Definition  \\ 
		\hline        
		$G, g$                          & Graph, subgraph.         \\
        $v, V, E$                       & Node (account), node set, edge set.         \\
        $\mathcal{N}(i)$                & 1-hop neighbor set of node $v_i$.         \\
        $\textbf{x}, \textbf{X}$        & Node feature, node feature matrix.         \\
        $\textbf{e}$                    & Edge feature.         \\
        $y, Y$                          & Account identity label, label set.         \\
        \hline
        $\textbf{h}$                    & Hidden representation of node feature.         \\
        $\textbf{s}, \textbf{g}$        & Subgraph representation before/after attentive pooling.         \\
        $\textbf{z}$                    & Projection representation of subgraph.         \\
        $f_\theta, f_\psi, f_\phi$      & Encoder, prediction head, projection head.         \\
        $\alpha, \beta$                 & Normalized attention scores.         \\
        $a$                             & Unnormalized attention scores.         \\
        $\Theta$                        & Weight parameters.         \\
        $T$                             & Graph augmentation method.         \\
        $\mathcal{L}$                   & Loss function.         \\
        \hline
		$N$                             & Parameter: batch size.              \\                  
        $h$                             & Parameter: hop in subgraph sampling.         \\
        $K$                             & Parameter: number of sampled neighbors per hop.         \\
        $\lambda$                       & Parameter: trade-off hyper-parameter in loss functions.         \\
		\hline\hline
	\end{tabular}
	}
\end{table}

\section{Related Work}
De-anonymization in blockchain has received considerable attention for market analysis, abnormal behavior detection, and law enforcement, deriving several mainstream techniques, such as address clustering and account identification.

\subsection{Address Clustering}
Early studies~\cite{meiklejohn2013fistful,spagnuolo2014bitiodine,reid2013analysis,androulaki2013evaluating,harrigan2016unreasonable,remy2017tracking,lischke2016analyzing} mainly focus on address clustering, also known as user re-identification or entity recognition.
Reid et al.~\cite{reid2013analysis} proposed the first heuristic for re-identification, named multi-input heuristic, which assumes that the input addresses of a particular transaction are possessed by the same entity. 
This heuristic is based on the fact that all private keys associated with addresses must be used conjointly to sign a transaction.
Androulaki et al.~\cite{androulaki2013evaluating} proposed the change address heuristic, which assumes that a new ``change'' address created by a transaction is likely controlled by the same entity that created the transaction, and has also been applied in~\cite{meiklejohn2013fistful,spagnuolo2014bitiodine}.
This heuristic stems from the change characteristics of Bitcoin that serves as a mechanism for enhancing user privacy.
Martin et al.~\cite{harrigan2016unreasonable} explored the reasons behind the effectiveness of using the multi-input heuristic for address clustering.
Cazabet et al.~\cite{remy2017tracking} proposed to construct an identity hint network and applied the Louvain algorithm~\cite{blondel2008fast} to detect communities representing the sets of addresses belonging to the same entities.

The aforementioned address clustering methods are widely used in Bitcoin.
Robin~\cite{klusman2018deanonymisation} analyzed the feasibility of two Bitcoin de-anonymization methods of IP linking and address clustering on Ethereum, and concluded that these two methods meet difficulties when applied to Ethereum due to technical differences.
Friedhelm~\cite{victor2020address} proposed three heuristics that exploit patterns related to deposit addresses, multiple participation in airdrops and token authorization mechanisms, and quantified the feasibility of each heuristic over the first four years of the Ethereum.
Shlomi et al.~\cite{linoy2021anonymizing} assumed that the smart contract code written by the same author has a unique style, and further linked contract addresses with similar code styles together, thinking that these addresses are generated by the same author.

\subsection{Account Identification}
Thanks to the openness of blockchain transactions, as well as the development of machine learning and network science, a new class of de-anonymization strategies --- account identification, has been proposed and comprehensively developed.
Existing account identification methods mainly concentrate on manual feature engineering and graph analytics.
\subsubsection{Manual Feature Engineering}
Manual feature engineering extremely relies on the prior knowledge of feature designers. 
Normally, the more expert experience involved, the more reliable the manual features are.
Toyoda et al.~\cite{toyoda2018multi} extracted seven statistical features such as the rate of bitcoin coinbase transactions to infer account identities.
Lin et al.~\cite{lin2019evaluation} designed various features associated with transaction timestamps and analyzed the importance of each one.
Bartoletti et al.~\cite{bartoletti2018data} designed the Gini coefficient and the characteristics of possible abnormal behavior patterns to infer the Ponzi accounts in the transaction network.
Marc et al.~\cite{jourdan2018characterizing} designed a large number of manual features associated with addresses, entities and graph motifs in Bitcoin transaction networks, and classified different Bitcoin entities via LightGBM~\cite{ke2017lightgbm}.
In addition, some emerging public blockchains contain smart contracts, providing new features. 
Huang et al.\cite{huang2020understanding} considered the calling information of smart contracts to expand the feature space, and realized the identification of bot accounts in EOSIO.
\subsubsection{Graph Analytics}
Massive transaction data can be modeled as graphs, and a considerable part of existing methods regards account identification as a classification task from a graph perspective.
Li et al.~\cite{li2020identifying} considered the topological features of accounts and found the difference in topological structure between the Ponzi accounts and the normal ones.
Yuan et al.~\cite{yuan2020detecting} applied graph random walks such as DeepWalk~\cite{perozzi2014deepwalk} and Node2vec~\cite{grover2016node2vec} to learn account features in the transaction graph.
Wu et al.~\cite{wu2020phishers} performed graph random walks by considering both the transaction amount and timestamp information, proposing a novel embedding method named Trans2Vec to extract the address feature for phishing detection.
Yuan et al.~\cite{yuan2020phishing} extracted the subgraphs for each target account and embedded their transaction topology via Graph2Vec~\cite{mlg201721}. 
Moreover, they introduced the SGN mechanism~\cite{xuan2019subgraph} to further enhance the transaction structure embedding. 
Chen et al.~\cite{chen2020phishing} also extracted transaction subgraphs and got the embeddings by a graph convolution layer combining graph auto-encoder in an unsupervised manner, finally achieving phishing detection by LightGBM.
Shen et al.\cite{shen2021identity} constructed the account interaction graphs using Ethereum and EOSIO data, and proposed an end-to-end graph convolution network model to identify different categories of accounts or bots.

\begin{table}[htp]
	\renewcommand\arraystretch{1.2}
	\centering
	\caption{Information of raw Ethereum block data.}
	\label{tb: raw-data}
	\resizebox{\linewidth}{!}{%
	\begin{tabular}{lcr} 
		\hline\hline
		\begin{tabular}[c]{@{}l@{}}Data~\\Field\end{tabular} & \begin{tabular}[c]{@{}l@{}}Custom \\Symbol\end{tabular} & Definition     \\
		\hline
		blockNumber     &                 & The block ID where the transcation is located.                             \\
		timestamp       & $d$             & The timestamp of a transaction.                                       \\
		from            & $v$             & The account that initiates the transaction.                             \\
		to              & $v$             & The account that receives the transaction.                              \\
		fromIsContract  &                 & Whether the transaction is sent by a CA.\\
		toIsContract    &                 & Whether the transaction is received by a CA.  \\
		callingFunction & $f$             & The name of function called if there is a contract call.                 \\
		value           & $w$             & The transaction amount.                                                 \\
		\hline\hline
	\end{tabular}}
  \end{table}

Besides the aforementioned methods, there are other frameworks to achieve identity identification. 
Phetsouvanh et al.\cite{phetsouvanh2018egret} proposed a graph mining technology to detect suspicious bitcoin flows and accounts by analyzing the path length and confluence account of the directed subgraph. 
Zhang et al.\cite{zhang2020anomaly} introduced the concept of meta-path from the heterogeneous network. This method deals with the bitcoin network from both static and dynamic perspectives and can effectively detect abnormal accounts and transactions.

\section{Account Interaction Graph Model}
\subsection{Problem Description}
In this paper, we mainly focus on identifying accounts in Ethereum via deep graph analytics, especially from a graph classification perspective.
A transaction graph constructed from blockchain transaction data is typically represented by a graph $G=(V, E, \textbf{X}, \textbf{E}, Y)$, where $V=\{v_1, v_2, \cdots, v_n\}$ is the set of account nodes, $E \subseteq \{(v_\textit{i},v_\textit{j})\mid v_\textit{i},v_\textit{j} \in V\}$ is the set of interaction edges, $\textbf{X} \in \mathbb{R}^{n\times F_1}$ is the node feature matrix, and $\textbf{E} \in \mathbb{R}^{m\times F_2}$ is the edge feature matrix (we assume, $|E| = m$).
We use $Y=\{(v_\textit{i}, y_\textit{i}) \mid v_\textit{i} \in V\}$ to represent the label set of partial account nodes.
The subgraph of an account node $v$ can be represented as $g_\textit{v} \subset G$.
For the given transaction graph $G$, subgraph-level account identification is to learn a function $f\left(g_\textit{v}\right) \mapsto y$ mapping the pattern of account subgraph $g_\textit{v}$ to the identity label $y$.

\subsection{Ethereum and Block Data}
Ethereum is the second-largest blockchain platform after Bitcoin, and it allows users to conduct complex transactions based on \textit{smart contracts}, which are applications that run on Ethereum virtual machines. 
An \textit{account} in Ethereum is an entity that owns Ether and can be divided into two categories: Externally Owned Account (EOA) and Contract Account (CA). 
EOA is controlled by a user who owns the private key of the account, and can initiate transactions. 
CA is controlled by smart contract code, which cannot initiate transactions actively and can only be executed according to the pre-written smart contract code after being triggered.
Between Ethereum accounts, there are usually two types of interactions: \textit{transaction} and \textit{contract call}.
The \textit{transaction} must be initiated by EOA, and can be received by EOA or CA. 
The \textit{contract call} refers to the process of triggering the smart contract code in CA to perform different operations.
The Ethereum blockchain is a succession of blocks, and each block contains a set of transactions and contract calls. 
The raw block data of Ethereum is structural and provides a wealth of information, as listed in Table~\ref{tb: raw-data}.

\begin{figure}[htp]
	\centering
		\includegraphics[width=\linewidth]{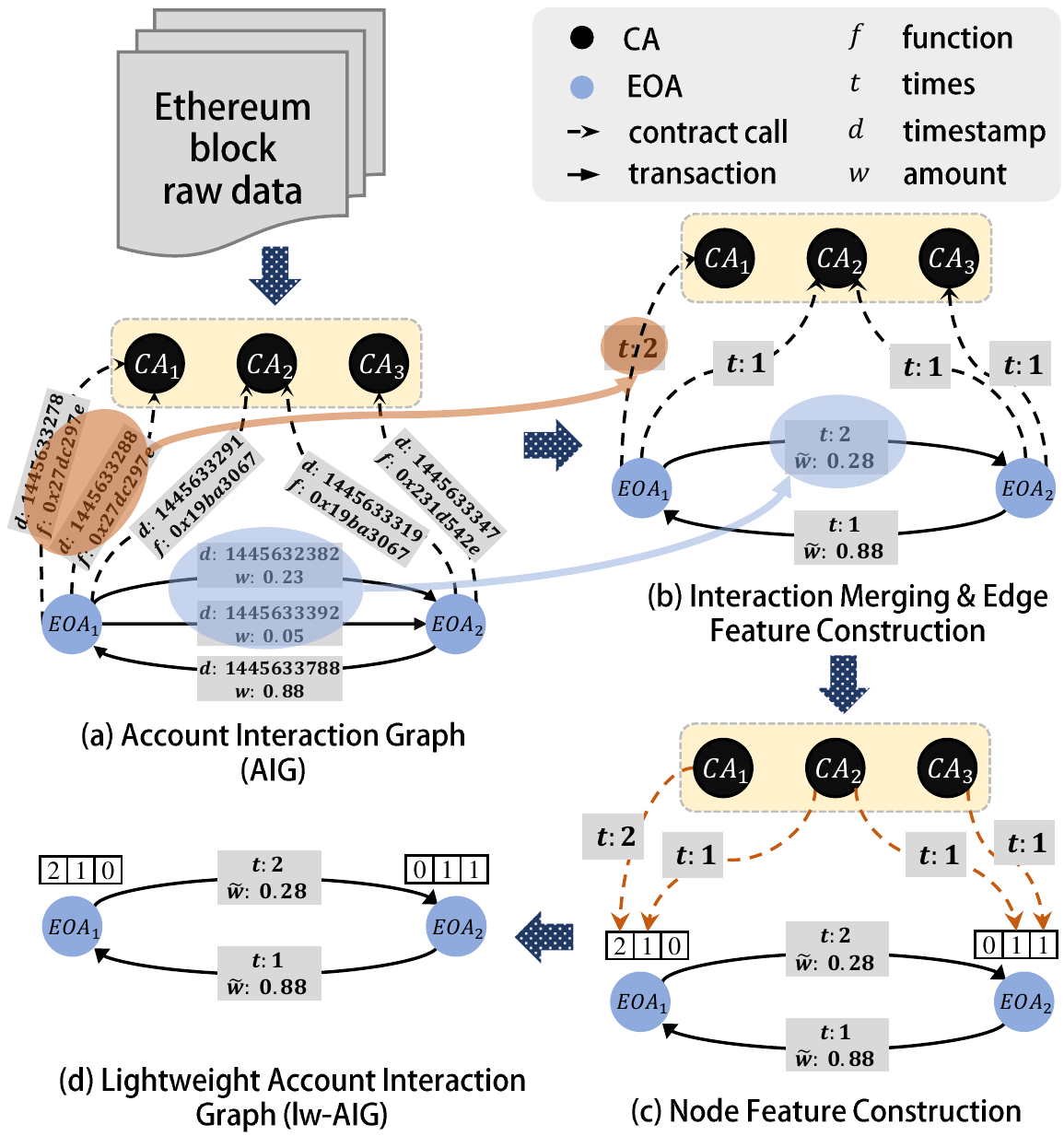}
		\caption{Constructing Account Interaction Graph and its lightweight version.}
		\label{fig: AIG}
\end{figure}

\begin{figure*}[htp]
	\centering
  \includegraphics[width=\textwidth]{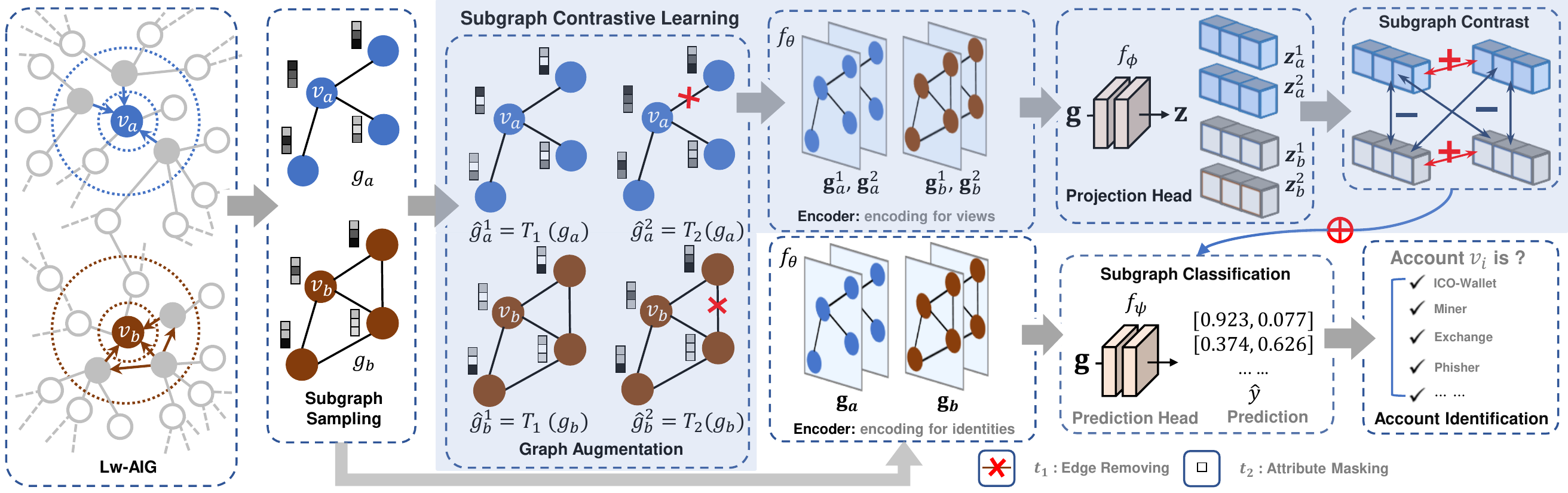}
  \caption{The architecture of Ethident. 
  The complete workflow proceeds as follows: 
  (1) sampling subgraphs centered on target accounts from lw-AIG; 
  (2) applying two augmentation operators on each subgraph to generate two correlated views;
  (3) encoding subgraphs and corresponding augmented views;
  (4) optimizing the GNN encoder by jointly training subgraph contrast and classification tasks.}
  \label{fig: framework}
\end{figure*}

\subsection{Account Interaction Graph}
The raw block data is informative and provides the details of transactions and contract calls, by which we can construct an Account Interaction Graph (AIG), as defined below.
\begin{definition}
	(\emph{Account Interaction Graph, AIG}): a directed, weighted and heterogeneous multigraph $G = \left(V_\textit{eoa}, V_\textit{ca}, E_\textit{t}, E_\textit{c}, Y\right)$, where $V_\textit{eoa}$ and $V_\textit{ca}$ are the set of EOA and CA respectively, $E_\textit{t}=\{(v_\textit{i}, v_\textit{j}, d, w)\mid v_\textit{i},v_\textit{j}\in V_\textit{eoa} \}$ is the directed edge set constructed from transaction information, and $E_\textit{c}=\{(v_\textit{i}, v_\textit{j}, d, f) \mid v_\textit{i}\in V_\textit{eoa}\cup V_\textit{ca}, v_\textit{j}\in V_\textit{ca}\}$ is the directed edge set constructed from contract call information. 
	The three edge attributes $d$, $w$, $f$ represent \textit{timestamp}, \textit{value} and \textit{callingFunction} respectively in Table~\ref{tb: raw-data}. 
	The AIG is partially labeled, i.e., a few EOA have identity labels $y$ and can compose the labeled node set $Y=\{(v_\textit{i}, y_\textit{i}) \mid v_\textit{i} \in V_\textit{eoa}\}$.
\end{definition}
The original AIG is a heterogeneous multigraph that has dense connections as well as different types of information attached to nodes and edges, as shown in Fig.~\ref{fig: AIG}(a).
The heterogeneity and multiple edges significantly increase the complexity of information mining.
So we further simplify the AIG into a homogeneous and more sparse graph by \textbf{interaction merging} and \textbf{feature construction}.

\begin{definition}
	(\emph{Lightweight Account Interaction Graph, lw-AIG}): a directed, weighted and homogeneous graph $G = (V_\textit{eoa}, \tilde{E}_\textit{t}, \mathbf{X}, \mathbf{E}, Y)$, where $\tilde{E}_\textit{t}=\{(v_\textit{i}, v_\textit{j}, t, \tilde{w} ) \mid v_\textit{i},v_\textit{j}\in V_\textit{eoa}\}$, $\mathbf{X}$ is the node feature matrix constructed from contract call information and $\mathbf{E}$ is the edge feature matrix.
	The edge attribute $t$ denotes the number of directed interactions from $v_\textit{i}$ to $v_\textit{j}$, and the edge attribute $\tilde{w}$ denotes total transaction amount from $v_\textit{i}$ to $v_\textit{j}$.
\end{definition}

\subsubsection{Interaction Merging and Edge Feature Construction}\label{sec: edge-feature}
During interaction merging, as shown in Fig.~\ref{fig: AIG}(b), multiple directed interactions (transactions or contract calls) from the source account $v_\textit{i}$ to the target account $v_\textit{j}$ will be merged into a single edge with a newly added edge attribute $t$ representing the number of merged interactions.
For transactions, another new edge attribute $\tilde{w}$ represents the total transaction amount of merged interactions.
In addition, a feature pruning operation will take effect, removing the two raw edge attributes of \textit{timestamp} $d$ and \textit{callingFunction} $f$.
Finally, we represent the edge feature vector for arbitrary transaction edge $(v_\textit{i},v_\textit{j})\in E_t$ as $\textbf{e}_\textit{ij}=[t,\tilde{w}]$.

\subsubsection{Node Feature Construction}
The behavior characteristics of an account are not only related to its transaction objects, amount and frequency, but also to the smart contracts it calls.
Accounts with different behavior patterns have different calling preferences for smart contracts.
Therefore, we can construct account features using the information on contract call, as shown in Fig.~\ref{fig: AIG}(c).
Specifically, let $n$ and $F$ be the number of EOA and CA respectively in AIG, we can construct an account feature matrix $\mathbf{X}\in\mathbb{R}^{n\times F}$ to represent the preference for contract call, as formulated below:
\begin{equation}
	\begin{array}{c}
		\mathbf{X}=[\mathbf{x}_{1} ; \cdots ; \mathbf{x}_\textit{i} \ ; \cdots ; \mathbf{x}_\textit{n}]^\top \vspace{1ex},  \\
		\mathbf{x}_\textit{i}=[t_{1}, \cdots , t_\textit{j} \  , \cdots , t_\textit{F}] \vspace{1ex},\\
		\text{where}\quad  t_\textit{j}=\left\{\begin{array}{l}
		t \quad \text {If there are}\  t \ \text{calls to } v_\textit{j}^\textit{ca}; \\
		0 \quad \text {If there is no call to } v_\textit{j}^\textit{ca};
		\end{array}\right.
	\end{array}
\end{equation}
Note that $\mathbf{x}_\textit{i}$ is the feature of $v_\textit{i}^\textit{eoa}$.
During feature construction, we convert the AIG to a homogeneous lw-AIG.

In summary, the node features of the lw-AIG reflect the contract call information, and the edge features reflect the transaction information.

\begin{figure}[htp]
	\centering
	\includegraphics[width=\linewidth]{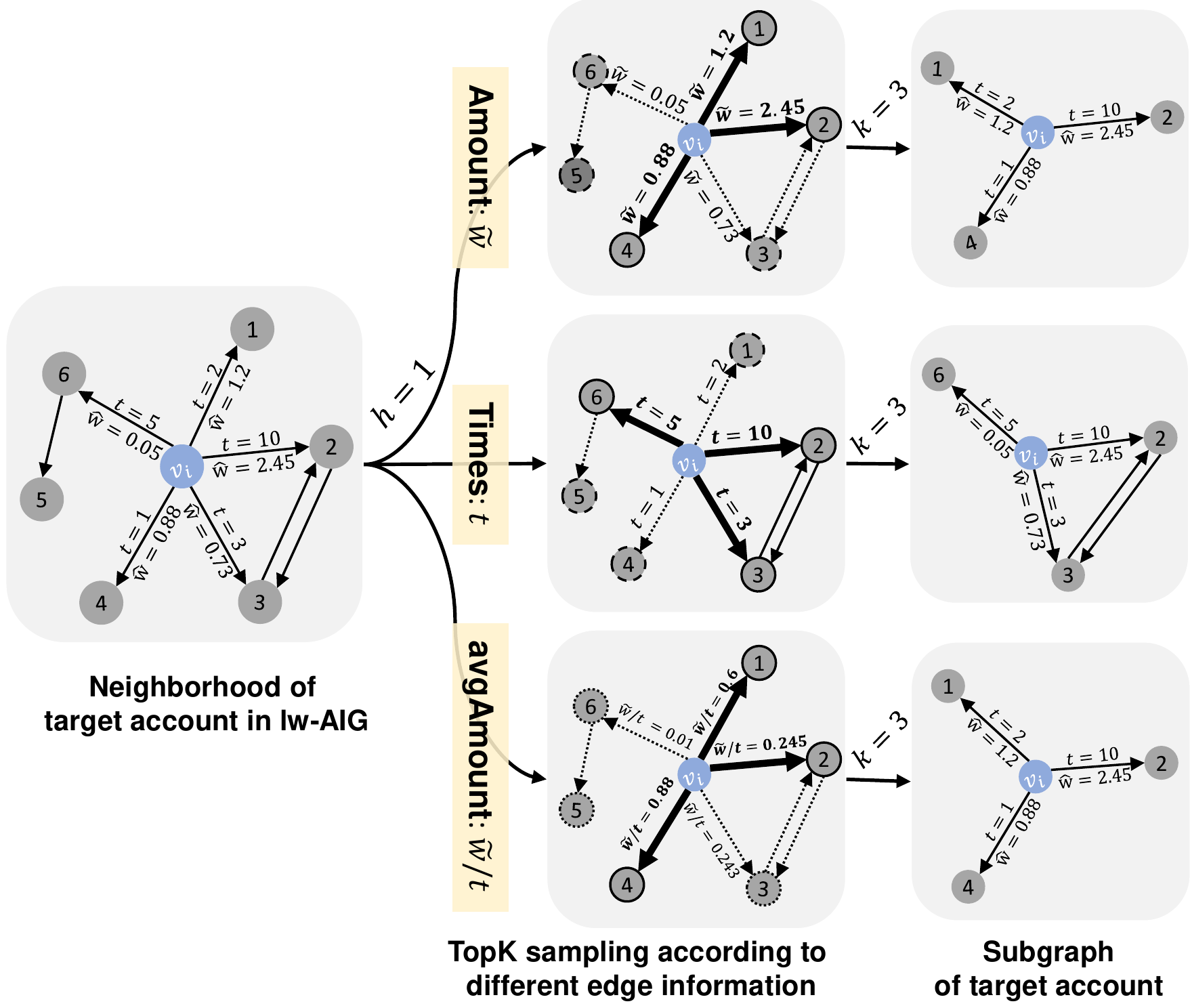}
	\caption{Subgraph sampling according to different edge information.}
	\label{fig: sample-1}
  \end{figure}

\section{Methodology}
In this section, we provide the details of the proposed framework \emph{Ethident}, as schematically depicted in Fig.~\ref{fig: framework}. 
For a target account $v_\textit{i}$, the input of \emph{Ethident} is the account interaction subgraph $g_\textit{i}$ sampled from lw-AIG, and the output is the predictive identity label $\hat{y}_\textit{i}$.
Our framework is mainly composed of the following components: 
(1) a subgraph extractor that captures the micro interaction subgraphs centered on target accounts from the lw-AIG topology; 
(2) a subgraph augmentation module that generates a series of variant graph views using various transformations on subgraphs; 
(3) a GNN encoder that encodes the subgraphs as expressive representations via hierarchical graph attention mechanism;
(4) a training module that jointly trains the subgraph contrast and classification tasks.
Next, we describe the details of each component.

\begin{figure*}[htp]
	\centering
  \includegraphics[width=\textwidth]{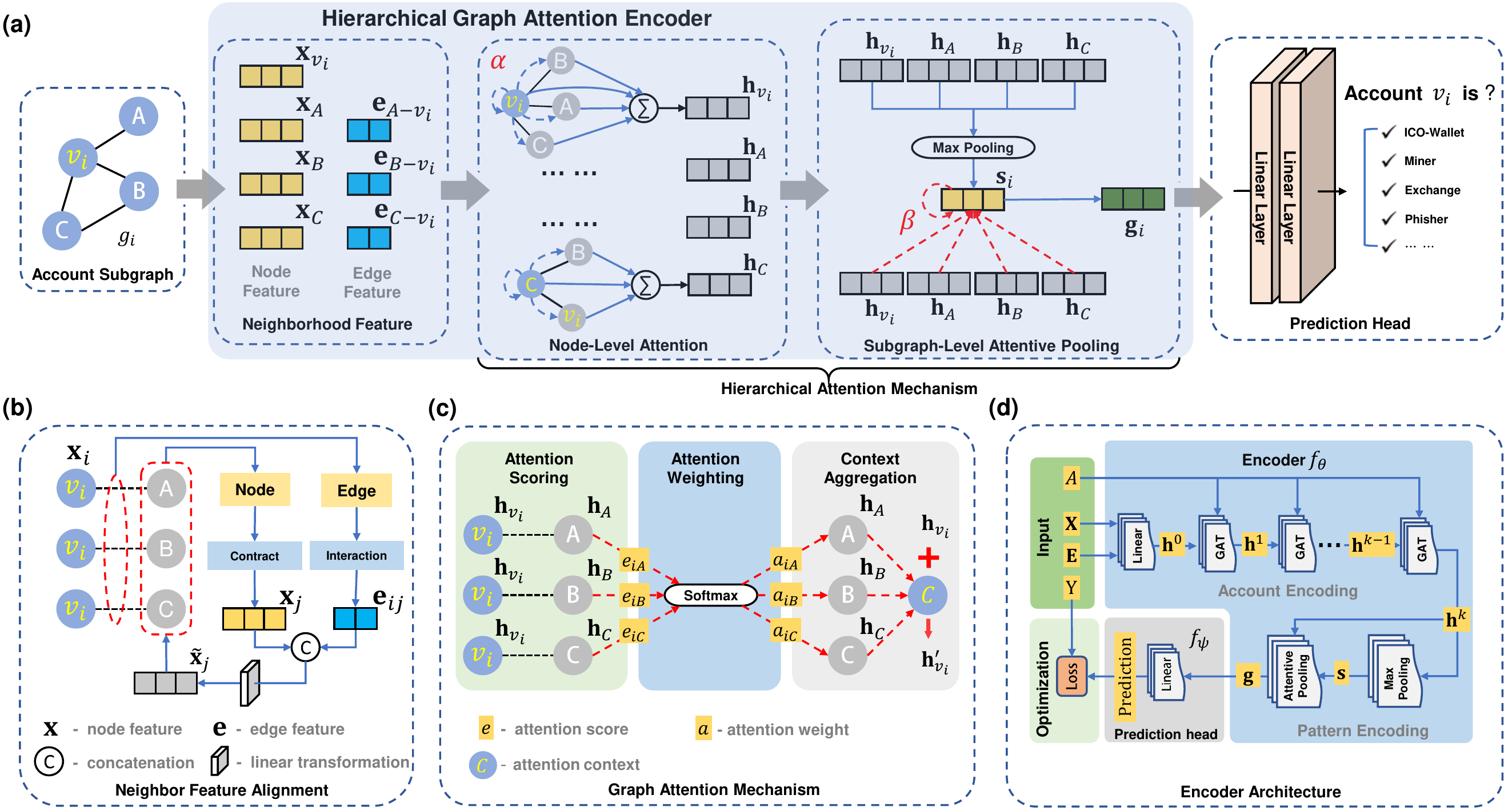}
  \caption{Schematic depiction of the hierarchical graph attention encoder (\emph{HGATE}): (a) the pipeline of \emph{HGATE} on account identification; (b) the process of neighbor feature alignment; (c) the illustration of graph attention mechanism; (d) the model architecture of \emph{HGATE}.} 
  \label{fig: encoder}
\end{figure*}

\subsection{Subgraph Sampling}\label{sec: sample}
The raw data contains tens of millions of blocks, making AIG a large-scale graph and not feasible for full-batch training of GNNs.
Even though the lightweight process greatly simplifies AIG, it still maintains a large number of EOA nodes.
On the other hand, existing account identification methods based on graph embedding or GNNs generally rely on full-batch training, which restricts their scalability on large-scale graphs for account representation learning.
Thus, we consider account identification as a subgraph-level classification task based on the following facts: 
(1) different types of accounts have different behavior patterns, implicit in their local structure;
(2) subgraph consisting of the target account and its local neighborhood information (neighbors and their interactions) is informative and plays a critical role in providing behavior patterns for account identification; 
(3) subgraph is the receptive field of the center target node, which is much smaller than the whole graph and allows for mini-batch training.

In this work, we consider subgraph sampling that allows for mini-batch training of GNNs on large-scale graphs.
We perform TopK sampling to obtain the $h$-hop interaction subgraphs according to different edge information: \textbf{Amount} ($\tilde{w}$), \textbf{Times} ($t$) or average Amount (\textbf{avgAmount}, $\tilde{w} / t$).
Specifically, for a target account node $v_\textit{i}$, we sample top-$K$ most important neighbors based on one of the edge attributes, and again sample top-$K$ most important neighbors for each account sampled at the previous hop, and recursive ones in the downstream hops.
The recursive sampling can be formulated as follows:
\begin{equation}
	V_k=\bigcup_{v \in V_{k-1}} \textit{topK}\left(\mathcal{N}_\textit{v}, K, \mathbf{E}\left[v, \mathcal{N}_\textit{v}, i\right]\right), \  i\in \{0,1,2\},
\end{equation}
where $V_k$ is the set of nodes sampled at hop $k$ and $V_0 = \{v_\textit{i}\}$, $\mathcal{N}_v$ is the 1-hop neighbor set of node $v$, $K$ is the number of sampled neighbors per hop, $\mathbf{E}\left[v, \mathcal{N}_\textit{v}, i\right]$ is the edge attributes of candidate interactions that guides the neighbor sampling, $i$ is an indicator of which edge attribute to use, and \textit{topK} is the function that returns the top-$K$ most important nodes.
After $h$ iterations, we obtain the account set $V_\textit{i}=\cup_{k=0}^{h} V_{k}$ sampled from lw-AIG, and the subgraph $g_\textit{i}$ of target account $v_\textit{i}$ can be induced by $V_\textit{i}$ from the lw-AIG.
Fig.~\ref{fig: sample-1} illustrates the process of subgraph sampling according to different edge information.
For the labeled target account set $Y$, their corresponding subgraphs form a dataset: $D = \{(g_\textit{i}, y_\textit{i})\mid \forall \left(v_\textit{i}, y_\textit{i}\right)\in Y \}$.
Note that we assign the label of the target account to the subgraph, and aim to learn a function mapping the subgraph patterns to account identity labels.

\subsection{Encoder Architecture}
The backbone of \emph{Ethident} is the designed GNN encoder named \emph{HGATE}, which is capable of learning expressive representations for accounts and their behavior patterns, as schematically depicted in Fig.~\ref{fig: encoder}.
This encoder learns account and pattern embeddings via a hierarchical attention mechanism, and can also implement account identification independently by following a prediction head, as shown in Fig.~\ref{fig: encoder}(a).
Next, we describe the details of our encoder $f_\theta $.

\subsubsection{Neighbor Feature Alignment}
For lw-AIG, its nodes and edges are encoded according to contract call and transaction information. 
Since our encoder is account-centric, each account $v_\textit{i}$ has its neighbor features that concatenate both neighboring account features ($\textbf{x}_\textit{j}$) and the connecting interaction features ($\textbf{e}_\textit{ij}$), represented as $[\mathbf{x}_\textit{j} \parallel \mathbf{e}_\textit{ij}]$.
Here we need to perform a column normalization on neighbor features to eliminate the dimensional differences between different attributes.
Note that the target account feature $\mathbf{x}_\textit{i} \in \mathbb{R}^{F}$ and its neighbor features $[\mathbf{x}_\textit{j} \parallel \mathbf{e}_\textit{ij}] \in \mathbb{R}^{F + 2}$ do not have the same dimension, so a linear transformation and a nonlinear activation are performed to align the feature dimension, as shown in Fig.~\ref{fig: encoder}(b).
This procedure can be achieved via a fully connected layer parameterized by $\mathbf{\Theta}_\textit{x}$ as follows, and generates aligned embeddings for neighbors of the target account.
\begin{equation}\label{eq:feature-align}
	\tilde{\mathbf{x}}_\textit{j} = \textit{LeakyRelu} \left(\mathbf{\Theta}_\textit{x} \cdot [\mathbf{x}_\textit{j} \parallel  \mathbf{e}_\textit{ij} ] \right).
\end{equation}

\subsubsection{Node-level Attention for Account Embedding}\label{sec: node-att}
This module aims to preserve the relevance of interactive accounts in the input subgraph, and learns account representation by focusing on the most relevant parts of the neighborhood.
When identifying a target account in the interaction subgraph, different neighboring accounts generally contribute differently to it.
For example, both accounts $v_\textit{a}$ and $v_\textit{b}$ have transactions with account $v_\textit{i}$, if $v_\textit{a}$ has many high-volume transactions with $v_\textit{i}$ while $v_\textit{b}$ has only one low-volume transaction with $v_\textit{i}$, or if $v_\textit{a}$ has a more similar preference of contract call with $v_\textit{i}$ than $v_\textit{b}$, then $v_\textit{a}$ often plays a more important role in identifying $v_\textit{i}$ since it preserves more information associated with the identity of $v_\textit{i}$.
Based on the above understanding and inspired by previous work~\cite{velickovic2018graph}, we utilize the node-level attention mechanism, as illustrated in Fig.~\ref{fig: encoder}(c), to learn the hidden representation of each account in the input subgraph by composing its neighbor features with different contributions (attentions).

Specifically, for arbitrary account $v_\textit{i}$ in the input subgraph $g$, the node-level attention mechanism learns the contribution attention scores for its neighbors $v_\textit{j}$, as follows:
\begin{equation}
	a_\textit{ij}^\textit{l} = \textit{LeakyRelu} \left(\mathbf{\Theta}_\textit{n}^\textit{l} \cdot [\textbf{h}_\textit{i}^\textit{l} \parallel  \textbf{h}_\textit{j}^\textit{l} ]\right),
\end{equation}
where a linear transformation parameterized by $\mathbf{\Theta}_\textit{n}^\textit{l}$ and a nonlinear \emph{LeakyRelu} activation are performed together to compute the importance of account $v_\textit{j}$'s hidden features to account $v_\textit{i}$ in $l$-th layer.
Subsequently, to make attention scores easily comparable across different accounts, the attention scores $a$ are further normalized using the softmax function over the neighbor accounts:
\begin{equation}
	\alpha_\textit{ij}^\textit{l}=\textit{Softmax} \left(a_\textit{ij}^\textit{l}\right)=\frac{\exp \left(a_\textit{ij}^\textit{l}\right)}{\sum_{x \in \mathcal{N}(i) \cup \{i\}} \exp \left(a_\textit{ix}^\textit{l}\right)},
\end{equation}
where $\mathcal{N}(i)$ is the 1-hop neighbor set of account $v_\textit{i}$.
Once obtained, the normalized attention scores are used to update the features of target account via neighborhood context aggregation:
\begin{equation}
	\mathbf{h}_\textit{i}^{\textit{l}+1}= \textit{Elu}\left(  \alpha_\textit{ii}^\textit{l} \cdot \Theta_\alpha^\textit{l} \cdot \mathbf{h}_\textit{i}^\textit{l}+\sum_{j \in \mathcal{N}(i)} \alpha_\textit{ij}^\textit{l} \cdot \Theta_\alpha^\textit{l} \cdot \mathbf{h}_\textit{j}^\textit{l} \right)  ,
\end{equation}
where a linear transformation parameterized by $\Theta_\alpha^\textit{l}$ and a nonlinear \emph{Elu} activation are used to compute the final output features.

The node-level attention mechanism serves for account embedding. 
Specifically, we use a stack of $k$ graph attention layers to capture the account features, as illustrated in Fig.~\ref{fig: encoder}(d).
The input of this stack is the initial account embedding $\mathbf{h}^0$ generated by a fully connected layer that accepts the account and interaction features (Eq.~(\ref{eq:feature-align})).
Notably, for a target account $v_\textit{i}$, its initial embedding is $\mathbf{h}_\textit{i}^0 = \mathbf{x}_\textit{i}$, and that of its neighboring accounts $v_\textit{j}$ is $\mathbf{h}_\textit{j}^0 = \tilde{\mathbf{x}}_\textit{j} $.
To better characterize accounts, the stack performs an iterative process of transferring, transforming, aggregating and updating the representation from interactive neighbors.
And after $k$ iterations, the final output account embeddings $\mathbf{h}^\textit{k}$ contain the interaction influence within $k$-hops.

\subsubsection{Subgraph-level Attentive Pooling for Pattern Embedding}
This module aims to characterize the behavior patterns of target accounts in the input subgraphs by extracting expressive subgraph-level features.
Actually, the behavior patterns of accounts are associated with their identities, i.e., accounts of different identities usually behave differently and have different subgraph patterns.
For ``Exchange'' subgraphs, the center node generally has an extremely high centrality and frequently interacts with surrounding neighbors, indicating high-volume transaction orders. 
For ``Ponzi'' or ``Gambling'' subgraphs, there exist two explicit characteristics indicating high investment and low return: (1) bi-directional edges (mutual transactions) between the center node and surrounding neighbors are rare, and the center node has high in-degree and low out-degree; (2) the incoming edges (investment) of the center node contain larger feature values associated with the digital currency than the outgoing edges (return).
Therefore, different accounts contribute differently to characterize the subgraph pattern reflecting the behavior of the target account.
Meanwhile, traditional practice usually captures the graph-level features using sum, mean or max pooling, resulting in feature smoothing and poor expressiveness.
Based on the above understanding, we design a novel subgraph-level attentive pooling module, as illustrated in Fig.~\ref{fig: encoder}(a), to learn the expressive representation of account subgraphs.

Specifically, for a subgraph $g$, we first obtain the initial subgraph-level embedding $\mathbf{s}$ by using global max pooling over all account embeddings in the subgraph:
\begin{equation}
	\mathbf{s} = \textit{MaxPooling}\left(\mathbf{h}^\textit{k}\right).
\end{equation}
Note that the input of the \emph{MaxPooling} layer is the final account embeddings $\mathbf{h}^\textit{k}$ generated in Sec.~\ref{sec: node-att}.
To better characterize the subgraph pattern, we update $\mathbf{s}$ by aggregating features of all accounts with different contributions (attentions).
In other words, for the initial subgraph embedding $\mathbf{s}$, we use an attention mechanism to learn the contribution attention score for arbitrary account $v_\textit{j}$ in the subgraph as follows:
\begin{equation}
	a_\textit{j} = \textit{LeakyRelu} \left(\mathbf{\Theta}_\textit{s} \cdot [\mathbf{s} \parallel  \mathbf{h}_\textit{j}^\textit{k} ]\right),
\end{equation}
where a linear transformation parameterized by $\mathbf{\Theta}_\textit{s}$ and a nonlinear \emph{LeakyRelu} activation are performed to compute the importance of account $v_\textit{j}$'s hidden features to the initial subgraph embedding $\mathbf{s}$.
Same as the node-level attention, a softmax function is applied to compute the normalized attention scores:
\begin{equation}
	\beta_\textit{j}=\textit{Softmax} \left(a_\textit{j}\right)=\frac{\exp (a_\textit{j})}{\sum_{x \in V_\textit{g} \cup \{s\}} \exp (a_\textit{x})},
\end{equation}
where $V_\textit{g}$ is the node set of subgraph $g$, and $a_\textit{s}$ is the self-attention score of $\mathbf{s}$.
Finally, the attentive pooling performs the update process as follows:
\begin{equation}
	\mathbf{g}= \textit{Elu}\left(  \beta_\textit{s} \cdot \Theta_\beta \cdot \mathbf{s}+\sum_{j \in V_\textit{g}} \beta_\textit{j} \cdot \Theta_\beta \cdot \mathbf{h}_\textit{j}^\textit{k} \right)  ,
\end{equation}
where a linear transformation parameterized by $\Theta_\beta$ and a nonlinear \emph{Elu} activation are used to compute the final subgraph embedding $\textbf{g}$ which characterizes the behavior pattern of the target account.

\subsection{Subgraph Contrastive Learning}
To alleviate the account label scarcity as well as learn highly-expressive pattern embeddings, our \emph{Ethident} introduce the contrastive self-supervision learning as a regularization to jointly train the GNN encoder.
\subsubsection{Graph Augmentation}
Contrastive learning relies heavily on well-designed data augmentation strategies for view generation.
So far, widely used techniques concentrate on structure-level and attribute-level augmentation~\cite{you2020graph, wang2020nodeaug, zhou2020data}.
In this paper, we use three categories of graph augmentation methods to generate the augmented views of subgraphs.

\begin{itemize}
	\item \textbf{Structure-level Augmentatio}n
		\iitem[$\circ$] \textbf{Node Dropping}: Each node has a certain probability $\mathcal{P} $ to be dropped from subgraph.
		\iitem[$\circ$] \textbf{Edge Removing}: Each edge has a certain probability $\mathcal{P} $ to be removed from subgraph.
	\item \textbf{Attribute-level Augmentation}
		\iitem[$\circ$] \textbf{Node Attribute Masking}: Each dimension of node features has a certain probability $\mathcal{P} $ to be set as zero.
		\iitem[$\circ$] \textbf{Edge Attribute Masking}: Each dimension of edge features has a certain probability $\mathcal{P} $ to be set as zero.
	\item \textbf{Sampling-based Augmentation} \\
	Since each subgraph is sampled from lw-AIG via one of the three sampling strategies mentioned in Sec.~\ref{sec: sample}, we can use the other two sampling methods to generate the sampling-based augmented views for this subgraph.
\end{itemize}

During graph augmentation, we generate two augmented views $\hat{g}_i^1$, $\hat{g}_i^2$ for each target account subgraph $g_\textit{i}$, and assign the identity label of target account to them as a pseudo label:
\begin{equation}
	\begin{array}{l}
		D_{\text{aug}1}=\left\{(\hat{g}_\textit{i}^{1}, y_\textit{i}) \mid  \hat{g}_\textit{i}^{1}=T_1(g_\textit{i}) ;(v_\textit{i}, y_\textit{i}) \in Y\right\} \vspace{1ex} ,\\
		D_{\text{aug}2}=\left\{(\hat{g}_\textit{i}^{2}, y_\textit{i}) \mid  \hat{g}_\textit{i}^{2}=T_2(g_\textit{i}) ;(v_\textit{i}, y_\textit{i}) \in Y\right\}.
	\end{array}
\end{equation}
In this way, we can scale up the training data and alleviate label scarcity.
The raw and augmented datasets will be used together to train the encoder.

\subsubsection{Subgraph Contrast}
In our contrastive learning setting, for each account subgraph $g_\textit{i}$, its two correlated views $\hat{g}_\textit{i}^1$ and $\hat{g}_\textit{i}^2$ are generated by undergoing two augmentation operators $T_1$ and $T_2$, where $\hat{g}_\textit{i}^1 = T_1(g_\textit{i})$ and $\hat{g}_\textit{i}^2 = T_2(g_\textit{i})$.
The correlated augmented views are fed into the encoder $f_\theta$, producing the whole subgraph representations $\mathbf{g}_\textit{i}^1$ and $\mathbf{g}_\textit{i}^2$.
Then they are mapped into an embedding space for contrast via a projection head $f_\phi $, yielding $\mathbf{z}_\textit{i}^1$ and $\mathbf{z}_\textit{i}^2$.
Note that $\theta$ and $\phi$ are the parameters of graph encoder and projection head respectively.
Finally, the goal of subgraph-level contrast is to maximize the consistency between two correlated augmented views of subgraphs in the contrastive space via minimizing the contrastive loss:
\begin{table}[htp]
	\centering
	\caption{Statistics of subgraph datasets sampled from account interaction graph of Ethereum.
	$|G|$ is the number of subgraphs, \textit{Avg.} $|V|$ and  \textit{Avg.} $|E|$ are the average number of nodes and edges in subgraphs respectively, $|\mathbf{x}|$ and $|\mathbf{e}|$ are the number of node and edge features in subgraphs.}
	\label{tb: subgraph}
	\resizebox{\linewidth}{!}{%
	\renewcommand\arraystretch{1.2}
	
	\begin{tabular}{lcccccc} 
	\hline\hline
	Dataset                  & $|G|$      & \textit{Avg.} $|N|$   & \textit{Avg.} $|E|$   & $|\mathbf{x}|$      & $|\mathbf{e}|$     \\ 
	\hline
	Eth-ICO-A                & 146        & 42.5                  & 141.3                 & 14885               & 2                  \\
	Eth-ICO-T                & 146        & 52.2                  & 152.6                 & 14885               & 2                  \\
	Eth-ICO-aA               & 146        & 42.4                  & 140.7                 & 14885               & 2                  \\
	
	Eth-Mining-A              & 130        & 23.7                  & 72.9                  & 14885               & 2                  \\
	Eth-Mining-T              & 130        & 24.7                  & 67.2                  & 14885               & 2                  \\
	Eth-Mining-aA             & 130        & 29.0                  & 91.9                  & 14885               & 2                  \\
	
	Eth-Exchange-A           & 386        & 33.6                  & 123.7                 & 14885               & 2                  \\
	Eth-Exchange-T           & 386        & 38.0                  & 113.4                 & 14885               & 2                  \\
	Eth-Exchange-aA          & 386        & 38.6                  & 148.6                 & 14885               & 2                  \\
	
	Eth-Phish\&Hack-A          & 5070       & 37.3                  & 110.8                 & 14885               & 2                  \\
	Eth-Phish\&Hack-T          & 5070       & 37.8                  & 101.6                 & 14885               & 2                  \\
	Eth-Phish\&Hack-aA         & 5070       & 37.8                  & 111.3                 & 14885               & 2                  \\
	\hline\hline
	\end{tabular}
	}
\end{table}
\begin{equation}\label{eq:loss-all}
  \mathcal{L}_\text{self} = \frac{1}{N} \sum_{i=1}^N  \mathcal{L}_\textit{i} ,
\end{equation}
where $N$ is the number of subgraphs in a batch (i.e., batch size). The loss for each subgraph can be computed as:
\begin{equation}\label{eq:loss-one}
  \mathcal{L}_\textit{i} = -\log \frac{e^{\textit{cos}(\mathbf{z}_\textit{i}^{1}, \mathbf{z}_\textit{i}^{2}) / \tau}}{\sum_{j=1, j \neq i}^{N} e^{\textit{cos}(\mathbf{z}_\textit{i}^{1}, \mathbf{z}_\textit{j}^{2}) / \tau}},
\end{equation}
where $\textit{cos}(\cdot, \cdot)$ is the cosine similarity function with $\textit{cos}(\mathbf{z}_\textit{i}^{1}, \mathbf{z}_\textit{j}^{2}) = {\mathbf{z}_\textit{i}^{1}}^\top  \mathbf{z}_\textit{j}^{2} / \| \mathbf{z}_\textit{i}^{1}\| \| \mathbf{z}_\textit{j}^{2}\|$, and $\tau$ is the temperature parameter.
The two correlated views $\mathbf{z}_\textit{i}^1$ and $\mathbf{z}_\textit{i}^2$ of account subgraph $g_\textit{i}$ are treated as a positive pair while the rest view pairs in the batch are treated as negative pairs.
The objective aims to maximize the consistency of positive pairs as opposed to negative ones, i.e., contrastive learning allows accounts of the same type to have more consistent representations, and makes accounts of different types have more obvious differences.

\subsection{Model Training}
We achieve account identification by a prediction head $f_\psi$, which maps the subgraph representations to labels reflecting account identity, yielding a classification loss:
\begin{equation}
	\mathcal{L}_{\text{pred}}=-\frac{1}{N} \sum_{i=1}^{N} y_\textit{i} \cdot \log \left(f_{\psi}\left(\mathbf{g}_\textit{i}\right)\right),
\end{equation}
where $\mathcal{L}_{\text{pred}}$ is the cross entropy loss.

The self-supervised subgraph contrast is a pretext task that serves as a regularization of the subgraph classification task.
The encoder \emph{HGATE} is jointly trained with the pretext and subgraph classification tasks.
The loss function consists of both the self-supervision and classification task loss functions, as formularized below:
\begin{equation}
	\mathcal{L} = \mathcal{L}_\text{pred}   + \lambda \cdot \mathcal{L}_\text{self},
\end{equation}
where $\lambda$ is a trade-off hyper-parameter controls the contribution of the self-supervision term.

\begin{table*}[htp]
	\renewcommand\arraystretch{1.2}
	\large
	\centering
	\caption{Summary of performance on account identification in terms of F1-score in percentage with standard deviation. 
	The highest performance is marked with boldface; the highest performance of different categories of baselines is underlined.
	The OPS. value stands for the optimal performance statistics of all graph embedding and GNN-based methods under different subgraph datasets.}
	\label{tb: result}
	\resizebox{\textwidth}{!}{%
	\begin{tabular}{l|ccc|ccc|ccc|ccc} 
	\hline\hline
	\multicolumn{1}{c|}{\multirow{3}{*}{Method}} & \multicolumn{12}{c}{Dataset (with different sampling strategy)}                                                                                                                                                                                                                                                                                                                                                                                                                                                                                            \\ 
	\cline{2-13}
	\multicolumn{1}{c|}{}       & \multicolumn{3}{c|}{Eth-ICO}                                                                                                                                                      & \multicolumn{3}{c|}{Eth-Mining}                                                                                                                                                        & \multicolumn{3}{c|}{Eth-Exchange}                                                                                                                                              & \multicolumn{3}{c}{Eth-Phish\&Hack}                                                                                   \\ 
	\cline{2-13}   
	\multicolumn{1}{c|}{}       & Amount                                             & Times                                                           & avgAmount                                                  & Amount                                                      & Times                                                            & avgAmount                                             & Amount                                              & Times                                                            & avgAmount                                             & Amount                                                 & Times                                                               & avgAmount                             \\ 
	\hline              
	Manual + LR                 & \multicolumn{1}{c}{\multirow{3}{*}{$\leftarrow$}}  & \multicolumn{1}{c}{76.73\footnotesize{$\pm$0.059}}              & \multicolumn{1}{c}{\multirow{3}{*}{$\rightarrow$}}         & \multicolumn{1}{c}{\multirow{3}{*}{$\leftarrow$}}           & \multicolumn{1}{c}{77.15\footnotesize{$\pm$0.036}}               & \multicolumn{1}{c}{\multirow{3}{*}{$\rightarrow$}}    & \multicolumn{1}{c}{\multirow{3}{*}{$\leftarrow$}}   & \multicolumn{1}{c}{87.34\footnotesize{$\pm$0.037}}               & \multicolumn{1}{c}{\multirow{3}{*}{$\rightarrow$}}    & \multicolumn{1}{c}{\multirow{3}{*}{$\leftarrow$}}     & \multicolumn{1}{c}{80.94\footnotesize{$\pm$0.042}}                  & \multicolumn{1}{c}{\multirow{3}{*}{$\rightarrow$}}                                                                        \\ 
	Manual + RF                 & \multicolumn{1}{c}{}                               & \multicolumn{1}{c}{\underline{79.52\footnotesize{$\pm$0.045}}}  & \multicolumn{1}{c}{}                                       & \multicolumn{1}{c}{}                                        & \multicolumn{1}{c}{81.32\footnotesize{$\pm$0.044}}               & \multicolumn{1}{c}{}                                  & \multicolumn{1}{c}{}                                & \multicolumn{1}{c}{90.13\footnotesize{$\pm$0.030}}               & \multicolumn{1}{c}{}                                  & \multicolumn{1}{c}{}                                  & \multicolumn{1}{c}{90.10\footnotesize{$\pm$0.007}}                  & \multicolumn{1}{c}{}                                                                                                      \\ 
	Manual + LGBM               & \multicolumn{1}{c}{}                               & \multicolumn{1}{c}{74.71\footnotesize{$\pm$0.046}}              & \multicolumn{1}{c}{}                                       & \multicolumn{1}{c}{}                                        & \multicolumn{1}{c}{\underline{82.16\footnotesize{$\pm$0.051}}}   & \multicolumn{1}{c}{}                                  & \multicolumn{1}{c}{}                                & \multicolumn{1}{c}{\underline{91.25\footnotesize{$\pm$0.030}}}   & \multicolumn{1}{c}{}                                  & \multicolumn{1}{c}{}                                  & \multicolumn{1}{c}{\underline{90.51\footnotesize{$\pm$0.007}}}      & \multicolumn{1}{c}{}                                                                                                                  \\           
	\hline                 
	DeepWalk + LR               & 56.69\footnotesize{$\pm$0.094}                     & 58.96\footnotesize{$\pm$0.054}                                  & 59.64\footnotesize{$\pm$0.049}                             & 56.94\footnotesize{$\pm$0.067}                              & 60.26\footnotesize{$\pm$0.066}                                   & 62.07\footnotesize{$\pm$0.084}                        & 59.17\footnotesize{$\pm$0.059}                      & 63.82\footnotesize{$\pm$0.049}                                   & 62.53\footnotesize{$\pm$0.045}                        & 58.73\footnotesize{$\pm$0.020}                        & 67.99\footnotesize{$\pm$0.028}                                      & 64.29\footnotesize{$\pm$0.011}           \\
	DeepWalk + RF               & 73.24\footnotesize{$\pm$0.078}                     & 68.93\footnotesize{$\pm$0.045}                                  & 67.12\footnotesize{$\pm$0.076}                             & 65.13\footnotesize{$\pm$0.045}                              & 71.58\footnotesize{$\pm$0.078}                                   & 65.40\footnotesize{$\pm$0.035}                        & 76.31\footnotesize{$\pm$0.049}                      & 80.53\footnotesize{$\pm$0.038}                                   & 80.19\footnotesize{$\pm$0.020}                        & 91.14\footnotesize{$\pm$0.012}                        & 89.77\footnotesize{$\pm$0.008}                                      & \underline{92.71\footnotesize{$\pm$0.008}}           \\
	DeepWalk + LGBM             & 58.28\footnotesize{$\pm$0.085}                     & 59.18\footnotesize{$\pm$0.076}                                  & 58.28\footnotesize{$\pm$0.061}                             & 60.29\footnotesize{$\pm$0.077}                              & 56.17\footnotesize{$\pm$0.078}                                   & 61.56\footnotesize{$\pm$0.062}                        & 73.73\footnotesize{$\pm$0.035}                      & 73.21\footnotesize{$\pm$0.062}                                   & 71.49\footnotesize{$\pm$0.040}                        & 89.78\footnotesize{$\pm$0.006}                        & 89.65\footnotesize{$\pm$0.009}                                      & 92.13\footnotesize{$\pm$0.007}           \\
	Node2Vec + LR               & 80.95\footnotesize{$\pm$0.054}                     & 62.36\footnotesize{$\pm$0.055}                                  & 79.37\footnotesize{$\pm$0.079}                             & 64.36\footnotesize{$\pm$0.061}                              & 72.85\footnotesize{$\pm$0.063}                                   & \underline{79.24\footnotesize{$\pm$0.046}}            & 66.06\footnotesize{$\pm$0.024}                      & 82.77\footnotesize{$\pm$0.029}                                   & 81.05\footnotesize{$\pm$0.029}                        & 66.53\footnotesize{$\pm$0.020}                        & 82.58\footnotesize{$\pm$0.007}                                      & 79.73\footnotesize{$\pm$0.009}           \\
	Node2Vec + RF               & \underline{88.21\footnotesize{$\pm$0.048}}         & \underline{78.91\footnotesize{$\pm$0.051}}                      & \underline{89.34\footnotesize{$\pm$0.042}}                 & \underline{74.37\footnotesize{$\pm$0.041}}                  & \underline{78.48\footnotesize{$\pm$0.068}}                       & 78.72\footnotesize{$\pm$0.058}                        & \underline{83.12\footnotesize{$\pm$0.039}}          & \underline{88.98\footnotesize{$\pm$0.033}}                       & 86.56\footnotesize{$\pm$0.012}                        & \underline{92.04\footnotesize{$\pm$0.003}}            & \underline{94.06\footnotesize{$\pm$0.005}}                          & 92.17\footnotesize{$\pm$0.007}          \\
	Node2Vec + LGBM             & 81.41\footnotesize{$\pm$0.064}                     & 65.53\footnotesize{$\pm$0.086}                                  & 80.95\footnotesize{$\pm$0.040}                             & 65.45\footnotesize{$\pm$0.073}                              & 73.36\footnotesize{$\pm$0.060}                                   & 78.72\footnotesize{$\pm$0.052}                        & 80.28\footnotesize{$\pm$0.033}                      & 87.68\footnotesize{$\pm$0.022}                                   & 85.79\footnotesize{$\pm$0.012}                        & 91.67\footnotesize{$\pm$0.006}                        & 94.00\footnotesize{$\pm$0.005}                                      & 91.90\footnotesize{$\pm$0.005}           \\
	Struc2Vec + LR              & 61.00\footnotesize{$\pm$0.051}                     & 58.96\footnotesize{$\pm$0.082}                                  & 55.10\footnotesize{$\pm$0.084}                             & 51.82\footnotesize{$\pm$0.059}                              & 61.56\footnotesize{$\pm$0.072}                                   & 59.80\footnotesize{$\pm$0.093}                        & 63.48\footnotesize{$\pm$0.038}                      & 57.02\footnotesize{$\pm$0.031}                                   & 59.09\footnotesize{$\pm$0.056}                        & 57.02\footnotesize{$\pm$0.012}                        & 59.47\footnotesize{$\pm$0.014}                                      & 54.89\footnotesize{$\pm$0.013}           \\
	Struc2Vec + RF              & 61.45\footnotesize{$\pm$0.069}                     & 60.32\footnotesize{$\pm$0.079}                                  & 60.09\footnotesize{$\pm$0.057}                             & 63.61\footnotesize{$\pm$0.080}                              & 69.44\footnotesize{$\pm$0.058}                                   & 60.81\footnotesize{$\pm$0.059}                        & 74.07\footnotesize{$\pm$0.030}                      & 71.66\footnotesize{$\pm$0.035}                                   & 69.60\footnotesize{$\pm$0.033}                        & 66.33\footnotesize{$\pm$0.016}                        & 68.93\footnotesize{$\pm$0.007}                                      & 65.02\footnotesize{$\pm$0.007}           \\
	Struc2Vec + LGBM            & 62.36\footnotesize{$\pm$0.053}                     & 56.01\footnotesize{$\pm$0.059}                                  & 58.05\footnotesize{$\pm$0.055}                             & 55.12\footnotesize{$\pm$0.041}                              & 60.25\footnotesize{$\pm$0.045}                                   & 62.80\footnotesize{$\pm$0.072}                        & 70.97\footnotesize{$\pm$0.030}                      & 71.49\footnotesize{$\pm$0.024}                                   & 68.13\footnotesize{$\pm$0.045}                        & 65.25\footnotesize{$\pm$0.018}                        & 68.53\footnotesize{$\pm$0.009}                                      & 64.52\footnotesize{$\pm$0.010}           \\ 
	Trans2Vec + LR              & 73.77\footnotesize{$\pm$0.081}                     & 61.05\footnotesize{$\pm$0.064}                                  & 59.64\footnotesize{$\pm$0.066}                             & 71.41\footnotesize{$\pm$0.068}                              & 53.78\footnotesize{$\pm$0.076}                                   & 75.87\footnotesize{$\pm$0.061}                        & 57.52\footnotesize{$\pm$0.034}                      & 76.81\footnotesize{$\pm$0.041}                                   & 79.75\footnotesize{$\pm$0.034}                        & 68.67\footnotesize{$\pm$0.019}                        & 63.05\footnotesize{$\pm$0.016}                                      & 58.21\footnotesize{$\pm$0.013}           \\
    Trans2Vec + RF              & 86.50\footnotesize{$\pm$0.051}                     & 71.02\footnotesize{$\pm$0.060}                                  & 73.16\footnotesize{$\pm$0.067}                             & 73.34\footnotesize{$\pm$0.072}                              & 61.85\footnotesize{$\pm$0.084}                                   & 77.87\footnotesize{$\pm$0.065}                        & 72.91\footnotesize{$\pm$0.046}                      & 82.65\footnotesize{$\pm$0.042}                                   & \underline{87.31\footnotesize{$\pm$0.025}}            & 89.94\footnotesize{$\pm$0.008}                        & 89.75\footnotesize{$\pm$0.006}                                      & 89.98\footnotesize{$\pm$0.007}           \\
    Trans2Vec + LGBM            & 73.91\footnotesize{$\pm$0.063}                     & 59.88\footnotesize{$\pm$0.059}                                  & 59.30\footnotesize{$\pm$0.059}                             & 67.94\footnotesize{$\pm$0.069}                              & 50.81\footnotesize{$\pm$0.084}                                   & 72.60\footnotesize{$\pm$0.072}                        & 68.19\footnotesize{$\pm$0.035}                      & 76.30\footnotesize{$\pm$0.040}                                   & 84.67\footnotesize{$\pm$0.030}                        & 87.90\footnotesize{$\pm$0.008}                        & 88.46\footnotesize{$\pm$0.010}                                      & 89.65\footnotesize{$\pm$0.006}            \\ 
	\hdashline                                          
	Graph2Vec + LR	            & 65.21\footnotesize{$\pm$0.061}	                 & 68.21\footnotesize{$\pm$0.067}	                               & 63.90\footnotesize{$\pm$0.063}	                            & 53.25\footnotesize{$\pm$0.068}	                          & 48.33\footnotesize{$\pm$0.080}	                                 & 56.24\footnotesize{$\pm$0.073}	                     & 66.45\footnotesize{$\pm$0.037}	                   & 61.58\footnotesize{$\pm$0.036}	                                  & 66.79\footnotesize{$\pm$0.043}	                      & 80.73\footnotesize{$\pm$0.009}	                      & 78.92\footnotesize{$\pm$0.006}	                                    & 79.94\footnotesize{$\pm$0.008}           \\
	Graph2Vec + RF	            & 66.71\footnotesize{$\pm$0.072}	                 & 71.15\footnotesize{$\pm$0.057}	                               & 65.75\footnotesize{$\pm$0.049}	                            & 53.68\footnotesize{$\pm$0.080}	                          & 49.94\footnotesize{$\pm$0.072}	                                 & 57.38\footnotesize{$\pm$0.081}	                     & 66.50\footnotesize{$\pm$0.044}	                   & 63.42\footnotesize{$\pm$0.033}	                                  & 64.72\footnotesize{$\pm$0.034}	                      & 80.91\footnotesize{$\pm$0.008}	                      & 78.97\footnotesize{$\pm$0.008}	                                    & 79.60\footnotesize{$\pm$0.008}           \\
	Graph2Vec + LGBM	        & 61.02\footnotesize{$\pm$0.070}	                 & 63.02\footnotesize{$\pm$0.063}	                               & 58.29\footnotesize{$\pm$0.071}	                            & 52.32\footnotesize{$\pm$0.072}	                          & 46.56\footnotesize{$\pm$0.089}	                                 & 57.62\footnotesize{$\pm$0.078}	                     & 64.09\footnotesize{$\pm$0.046}	                   & 60.65\footnotesize{$\pm$0.041}	                                  & 61.17\footnotesize{$\pm$0.041}	                      & 82.11\footnotesize{$\pm$0.009}	                      & 80.43\footnotesize{$\pm$0.009}	                                    & 81.44\footnotesize{$\pm$0.006}           \\
	\hline                                                                                                                               
	GCN                         & 87.57\footnotesize{$\pm$0.112}                     & 86.73\footnotesize{$\pm$0.109}                                  & 86.89\footnotesize{$\pm$0.126}                             & 75.25\footnotesize{$\pm$0.130}                              & \underline{82.91\footnotesize{$\pm$0.098}}                       & 83.52\footnotesize{$\pm$0.073}                        & 90.23\footnotesize{$\pm$0.021}                      & 89.85\footnotesize{$\pm$0.026}                                   & 89.79\footnotesize{$\pm$0.026}                        & \underline{96.49\footnotesize{$\pm$0.005}}            & \underline{96.15\footnotesize{$\pm$0.006}}                          & \underline{96.49\footnotesize{$\pm$0.006}}           \\
	GAT                         & 88.44\footnotesize{$\pm$0.084}                     & 90.02\footnotesize{$\pm$0.047}                                  & 86.11\footnotesize{$\pm$0.147}                             & 80.28\footnotesize{$\pm$0.111}                              & 78.63\footnotesize{$\pm$0.152}                                   & 82.71\footnotesize{$\pm$0.084}                        & \underline{91.01\footnotesize{$\pm$0.024}}          & 91.42\footnotesize{$\pm$0.023}                                   & \underline{90.70\footnotesize{$\pm$0.021}}            & 95.87\footnotesize{$\pm$0.006}                        & 95.68\footnotesize{$\pm$0.005}                                      & 95.91\footnotesize{$\pm$0.005}           \\
	GIN                         & 75.20\footnotesize{$\pm$0.104}                     & 78.66\footnotesize{$\pm$0.102}                                  & 75.17\footnotesize{$\pm$0.096}                             & 59.07\footnotesize{$\pm$0.106}                              & 57.48\footnotesize{$\pm$0.113}                                   & 63.99\footnotesize{$\pm$0.129}                        & 81.56\footnotesize{$\pm$0.074}                      & 84.35\footnotesize{$\pm$0.067}                                   & 85.42\footnotesize{$\pm$0.055}                        & 95.89\footnotesize{$\pm$0.007}                        & 95.79\footnotesize{$\pm$0.007}                                      & 95.88\footnotesize{$\pm$0.005}           \\
	$\text{I}^2\text{BGNN}$-A   & \underline{92.13\footnotesize{$\pm$0.039}}         & 91.10\footnotesize{$\pm$0.042}                                  & \underline{92.41\footnotesize{$\pm$0.033}}                 & 82.24\footnotesize{$\pm$0.056}                              & 79.52\footnotesize{$\pm$0.124}                                   & 80.20\footnotesize{$\pm$0.117}                        & 89.64\footnotesize{$\pm$0.021}                      & 91.63\footnotesize{$\pm$0.021}                                   & 88.76\footnotesize{$\pm$0.032}                        & 95.94\footnotesize{$\pm$0.005}                        & 95.93\footnotesize{$\pm$0.005}                                      & 96.07\footnotesize{$\pm$0.004}           \\
	$\text{I}^2\text{BGNN}$-T   & 91.02\footnotesize{$\pm$0.084}                     & \underline{91.72\footnotesize{$\pm$0.034}}                      & 90.13\footnotesize{$\pm$0.085}                             & \underline{82.64\footnotesize{$\pm$0.077}}                  & 82.13\footnotesize{$\pm$0.101}                                   & \underline{83.84\footnotesize{$\pm$0.062}}            & 89.28\footnotesize{$\pm$0.025}                      & \underline{91.87\footnotesize{$\pm$0.022}}                       & 89.87\footnotesize{$\pm$0.028}                        & 95.99\footnotesize{$\pm$0.004}                        & 95.98\footnotesize{$\pm$0.005}                                      & 96.11\footnotesize{$\pm$0.004}           \\ 
	\hdashline                                                                                                   
	\textbf{Ethident} (w/o GC)  & 93.02\footnotesize{$\pm$0.029}                     & \textbf{93.36}\textbf{\footnotesize{$\pm$0.032}}                & \textbf{94.38}\textbf{\footnotesize{$\pm$0.028}}           & 85.62\footnotesize{$\pm$0.060}                              & 83.68\footnotesize{$\pm$0.080}                                   & 84.91\footnotesize{$\pm$0.051}                        & 92.28\footnotesize{$\pm$0.027}                      & 92.77\footnotesize{$\pm$0.027}                                   & 92.39\footnotesize{$\pm$0.021}                        & 97.79\footnotesize{$\pm$0.003}                        & 97.37\footnotesize{$\pm$0.004}                                      & 97.80\footnotesize{$\pm$0.003}   \\
	\textbf{Ethident}           & \textbf{94.05}\textbf{\footnotesize{$\pm$0.034}}   & 92.76\footnotesize{$\pm$0.038}                                  & 94.05\footnotesize{$\pm$0.033}                             & \textbf{86.38}\textbf{\footnotesize{$\pm$0.049}}            & \textbf{87.00}\textbf{\footnotesize{$\pm$0.040}}                 & \textbf{85.30}\textbf{\footnotesize{$\pm$0.057}}      & \textbf{93.16}\textbf{\footnotesize{$\pm$0.021}}    & \textbf{93.55}\textbf{\footnotesize{$\pm$0.027}}                 & \textbf{93.34}\textbf{\footnotesize{$\pm$0.022}}      & \textbf{97.93}\textbf{\footnotesize{$\pm$0.002}}      & \textbf{97.58}\textbf{\footnotesize{$\pm$0.004}}                    & \textbf{97.98}\textbf{\footnotesize{$\pm$0.003}}  \\
	\hline                                                                                   
	\emph{OPS.}                 & \textbf{11}                                        & 7                                                               & 5                                                          & 2                                                           & 4                                                                & \textbf{16}                                           & 6                                                   & \textbf{11}                                                      & 5                                                     & 6                                                     & 7                                                                   & \textbf{10}                                          \\                                                                                                                                                                                  
	
	\hline\hline
	\end{tabular}}
\end{table*}

\section{Experiments}
\subsection{Data Preparation}\label{sec: data}
We intercept the first 10 million block data (the time interval is between ``2015-07-03'' to ``2020-05-04'') from the Xblock website\footnote{http://xblock.pro/}~\cite{zhen2020xblock}.
Within this time interval, we can extract in total 309,010,831 transactions and 175,351,541 contract calls, involving 90,193,755 EOA and 16,221,914 CA.
Account identity labels are obtained from Label Word Cloud in Ethereum blockchain browser\footnote{https://etherscan.io/labelcloud}, including 73 \emph{ICO-wallet}, 65 \emph{Mining}, 193 \emph{Exchange} and 2,535 \emph{Phish/Hack}.

These four types of accounts are prevalent on blockchain platforms, and have received widespread attention. 
It is of sufficient practical significance to identify whether an account belongs to these types, especially for phish and hack accounts.
For each type of identity label (\emph{ICO-wallet}, \emph{Mining}, \emph{Exchange} and \emph{Phish/Hack}), we sample all target account subgraphs with this label as the positive sample, as well as the same number of randomly sampled account subgraphs with other labels as the negative sample.
We perform subgraph sampling for each labeled account according to different edge information (\textbf{Amount}, \textbf{Times} or \textbf{avgAmount}), yielding three types of datasets whose names are suffixed with ``-A'', ``-T'', ``-aA'', respectively.
Table~\ref{tb: subgraph} shows the specifications of subgraph datasets sampled from lw-AIG with $h=2$ and $K=20$ .

\subsection{Comparison Methods}
To illustrate the effectiveness of our \emph{Ethident} on account identification, we compare with three broad categories of methods: manual feature engineering, graph embedding methods and GNN-based methods.

For manual feature engineering that is the most common and simplest method for account identification, we design 16 manual features for Ethereum accounts according to the prior knowledge and the characteristics of raw Ethereum data, as detailedly described in Appendix~\ref{app: manual}, yielding account embeddings with dimension size of 16.
For graph embedding methods, we consider DeepWalk~\cite{perozzi2014deepwalk}, Node2Vec~\cite{grover2016node2vec}, Struc2Vec~\cite{ribeiro2017struc2vec}, Trans2Vec~\cite{wu2020phishers} and Graph2Vec~\cite{mlg201721} for account embedding.
For the above two categories of methods, we achieve account identification by feeding the generated account embeddings into three kinds of machine learning classifiers: Logistic Regression (LR), Random Forest (RF) and LightGBM (LGBM).

For GNN-based methods, we first compare with three commonly used GNNs: GCN~\cite{kipf2017semi}, GAT~\cite{velickovic2018graph}, and GIN~\cite{xu2018powerful}, which are adjusted for subgraph classification by following with a pooling layer and a prediction head.
We also compare with previous related work for account identification: $\text{I}^2\text{BGNN}$, which achieves account identification based on different edge information, yielding two variants: $\text{I}^2\text{BGNN}$-A and $\text{I}^2\text{BGNN}$-T.

\begin{figure*}[htp]
	\centering
  \includegraphics[width=\textwidth]{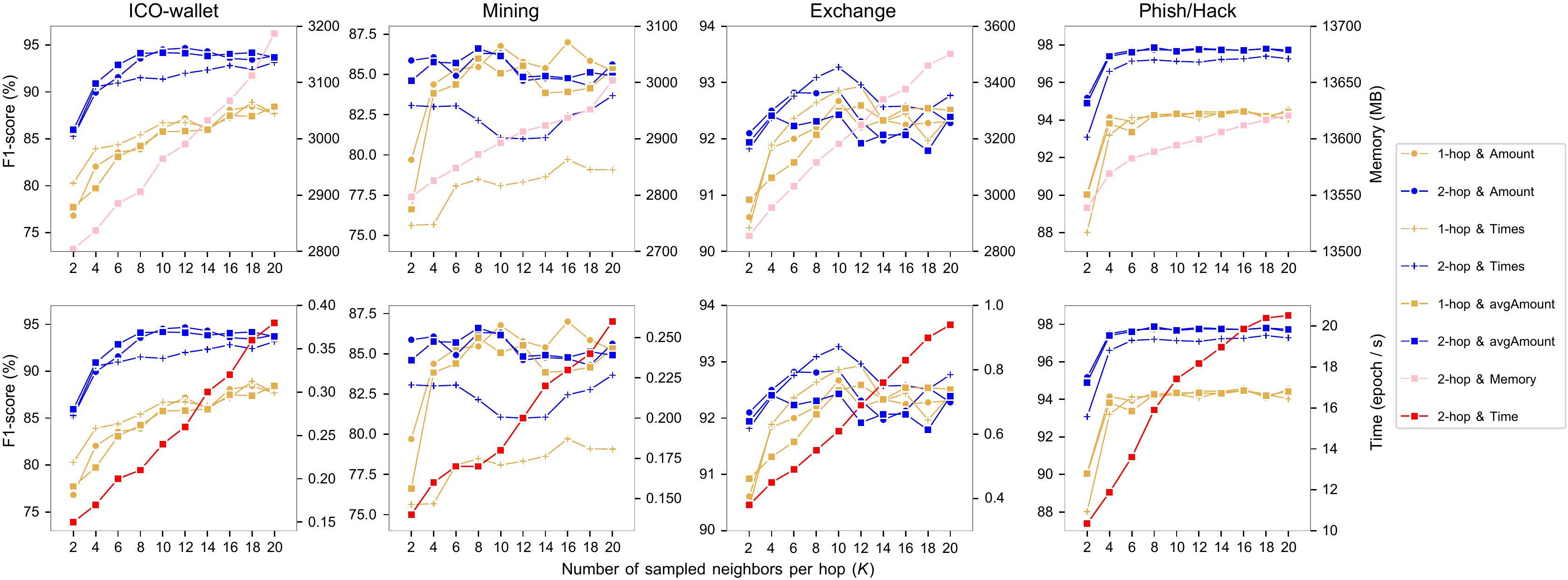}
  \caption{Impact of sampling scale on performance and consumption (memory and time).}
  \label{fig: line-size}
\end{figure*}

\subsection{Experimental Settings} \label{sec: exp}
For subgraph sampling in \emph{Ethident}, we set the subgraph hop $h$ to 2 and sample $K=20$ neighbors per hop.
For graph augmentation, we set the probability $\mathcal{P} $ to 10\%.
For the encoder \emph{HGATE}, we stack $k=2$ graph attention layers with the hidden dimension of 128 for account embedding, and use global max pooling for initial subgraph embedding.
In addition, the projection head $f_\phi$ is a two-layer perceptron with Relu activation and linear skip connection, and the prediction head $f_\psi$ is a two-layer perceptron with Relu and Softmax activation.
We set the temperature parameter $\tau$ and trade-off coefficient $\lambda$ to 0.2 and 0.01, respectively.

For GCN, GAT, GIN and $\text{I}^2\text{BGNN}$, the number of the corresponding message passing layers are 2, 2, 5 and 2 respectively.
The global max pooling is used for final subgraph embedding.
For all GNN-based methods, we set the embedding dimension, batch size $N$, learning rate, dropout to 128, 32, 0.001, 0.2, respectively.
During model training, we use early stopping with patience of 20.

For DeepWalk, Node2Vec, Struc2Vec and Trans2Vec, we set the length of walks to 20, the number of walks to 40, and the context size to 3.
For Node2Vec, we set the return parameter $p$ and in-out parameter $q$ to 0.25 and 0.4, respectively.
For the above four random walk-based methods which are extremely inefficient on large-scale graphs, we generate a training graph by sampling the connected subgraph containing all target accounts and their partial 2-hop neighbors from lw-AIG.
For Graph2Vec, we set the number of Weisfeiler-Lehman iterations to 2, the downsampling frequency to 0.0001, the minimal count of graph feature occurrences to 5, the epoch to 500 and the learning rate to 0.025.
For the above graph embedding methods, we set the dimension of output account embedding to 128.

For each subgraph dataset sampled from lw-AIG, we split it into training, validation and testing sets with a proportion of 1:1:1, repeat 3-fold cross validation 10 times and report the average micro-F1 score as well as standard deviation.

\subsection{Evaluation on Account Identification}
We evaluate our \emph{Ethident} on account identification and the results are presented in Table~\ref{tb: result}, from which we can observe that our \emph{Ethident} achieves state-of-the-art results with respect to comparison methods.
Specifically, our \emph{Ethident} significantly outperforms manual feature engineering and graph embedding methods across all datasets, and yields 2.09\% $\sim$ 18.27\% relative improvement over best baselines in terms of F1-score, indicating that the learned subgraph features are better at capturing the behavior patterns of accounts than manual or shallow topology features.
When compared to GNN-based methods, our \emph{Ethident} surpasses strong baselines: we observe 1.13\% $\sim$ 4.93\% relative improvement over best baselines.

These observations meet our intuition.
As we can see, the performance of manual features and graph embedding methods varies largely across different datasets with comparatively lower performance rankings. 
This is consistent with our assertion that they have limited expressiveness for different kinds of account subgraphs.
Because manual features and graph embedding methods cannot learn task-related features in an end-to-end manner, they rely heavily on the choice of classifiers to achieve relatively high performance. 
Meanwhile, classic GNN baselines normally surpass the manual features and graph embedding methods since they learn simultaneously from both graph topology and latent features.
Nevertheless, these baselines like GCN and GAT disregard the important edge information and generate subgraph-level features via naive pooling operations.
The two variants of $\text{I}^2\text{BGNN}$ only consider one single interaction information and disregard others.
Combining with the above analysis, we know that our \emph{Ethident} learns from both node and edge information associated with behavior patterns and identities of accounts, and uses a hierarchical attention mechanism to effectively characterize node-level account features and subgraph-level behavior patterns, reasonably achieving superior performance on account identification.

\begin{figure*}[htp]
	\centering
  \includegraphics[width=\textwidth]{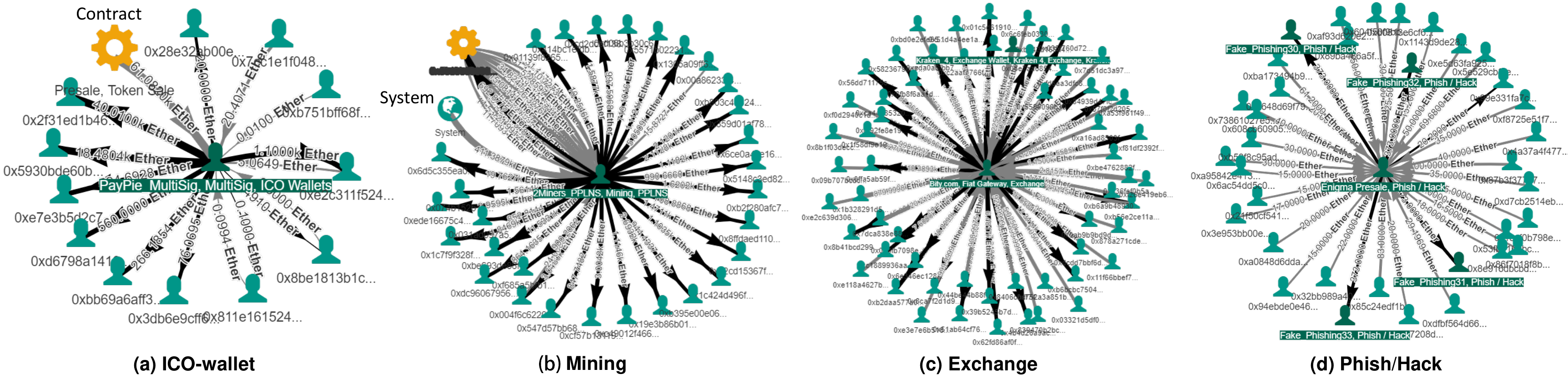}
  \caption{Different categories of accounts generally have different behavior patterns, as embodied in their micro interaction subgraphs.
  Here we present some interesting accounts to help explanation.}
  \label{fig: pattern}
\end{figure*}

\subsection{Pattern Analysis in Micro Interaction Subgraphs}
After evaluating the overall performance of our method, we investigate the behavior patterns of different accounts using experimental results on micro interaction subgraphs.
Furthermore, we list the following \textbf{Obs}ervations as well as explainable analysis.

\subsubsection{\textbf{Obs. 1. Larger subgraphs generally contain more critical identity-related pattern information}}
We analyze the impact of subgraph scale by evaluating our encoder on subgraph datasets with different scale settings.
Specifically, we vary $h$ in $\{1,2\}$ and $K$ in $\{2,4, \cdots , 20 \}$.
We observe that it is generally better to infer from 2-hop subgraphs than 1-hop ones, especially for \emph{ICO-wallet} and \emph{Phish/Hack}, judging from Fig.~\ref{fig: line-size}.
Meanwhile, as the size of subgraphs increases, the performance becomes better first and then remains stable or fluctuates slightly in most cases.
The above phenomenon suggests that subgraphs with larger scale benefit account identification more, meeting our intuition that larger subgraphs generally contain more critical pattern information associated with account identities.

\subsubsection{\textbf{Obs. 2. Different subgraph information highlights the behavior patterns of accounts with different contributions}}
For each category of accounts, the performance of all methods except manual features varies largely across the datasets with different sampling strategies.
And we count the number of optimal performances obtained by all methods under different sampling strategies, yielding the \emph{OPS.} values.
Judging from the bottom row in Table~\ref{tb: result}, we have reasonable explanations for such phenomenon that different sampling strategies benefit differently for account identification.

As we know, different categories of accounts have different behavior patterns that are embodied in their micro interaction subgraphs.
Here we present some interaction subgraphs of real accounts to help explain, as shown in Fig.~\ref{fig: pattern}.
Note that only the cumulative transaction amount of Ether is displayed between any two connected nodes in the interaction subgraphs.

\begin{itemize}
	\item \emph{\textbf{ICO-wallet}}: Initial Coin Offering (ICO) is a financing method that raises funds for blockchain projects by issuing tokens. 
	ICO projects usually pre-sell tokens in exchange for a large amount of Ether, and after a period, the projects will give supporters a certain return on their investment.
	The key behavior pattern is represented as a large number of outgoing edges with a certain \textbf{Amount} of investment rewards from the center ICO account to the surrounding supporters.
	Since investment actions generally involve a higher transaction amount, sampling interaction subgraphs according to \textbf{Amount} information can maximally preserve the behavior pattern of ICO accounts.
	\item \emph{\textbf{Mining}}: Mining pooling is a cooperative mining team that shares computational power to find blocks. 
	The mining pool will receive a large amount of mining rewards issued by the system, and distribute them to subordinate miners according to the proof-of-work (PoW) consensus protocol.
	The key behavior pattern is represented as a large number of outgoing edges with a certain amount of cumulative rewards from the center mining pool to the surrounding miner nodes.
	Since the block reward of Ethereum is fixed for a period, miners in the same mining pool generally have a relatively stable average mining income, which inspires us to use the average amount (\textbf{avgAmount}) information to guide the sampling of interactive subgraphs.
	\item \emph{\textbf{Exchange}}: The exchange is a platform that provides users with asset transaction matching and clearing services.
	Exchange accounts usually interact frequently with their clients to process a large number of transaction orders, and behave as hub nodes with extremely high centrality (i.e., large in-degree and out-degree) in the interaction graphs.
	So sampling interaction subgraphs according to \textbf{Times} information benefits more.
	\item \emph{\textbf{Phish/Hack}}: Both \emph{Phishers} and \emph{Hackers} engage in illegal fraud activities, in which they usually spread a large number of websites, emails or links containing viruses, Trojans, unwanted software, etc., and trick the recipient into doing remittances directly or providing the sensitive information of system privileges.
	As shown in Fig.~\ref{fig: pattern}(d), the center phish/hack account receives large amounts of Ether through various scams and disperses them to other phish/hack accounts for concealment.
	The key behavior pattern has one explicit characteristic: bi-directional edges (mutual transactions) between the center node and surrounding ones are rare, and the center node has high in-degree and low out-degree. 
	Since illegal frauds such as fake token exchange or ransomware often set a specific threshold amount or fixed ransom which can be reflected in the \textbf{average amount} information, using subgraphs sampled according to \textbf{avgAmount} information may benefit more for identifying \emph{Phish/Hack} accounts.

\end{itemize}

\begin{figure*}[htp]
	\centering
  \includegraphics[width=\textwidth]{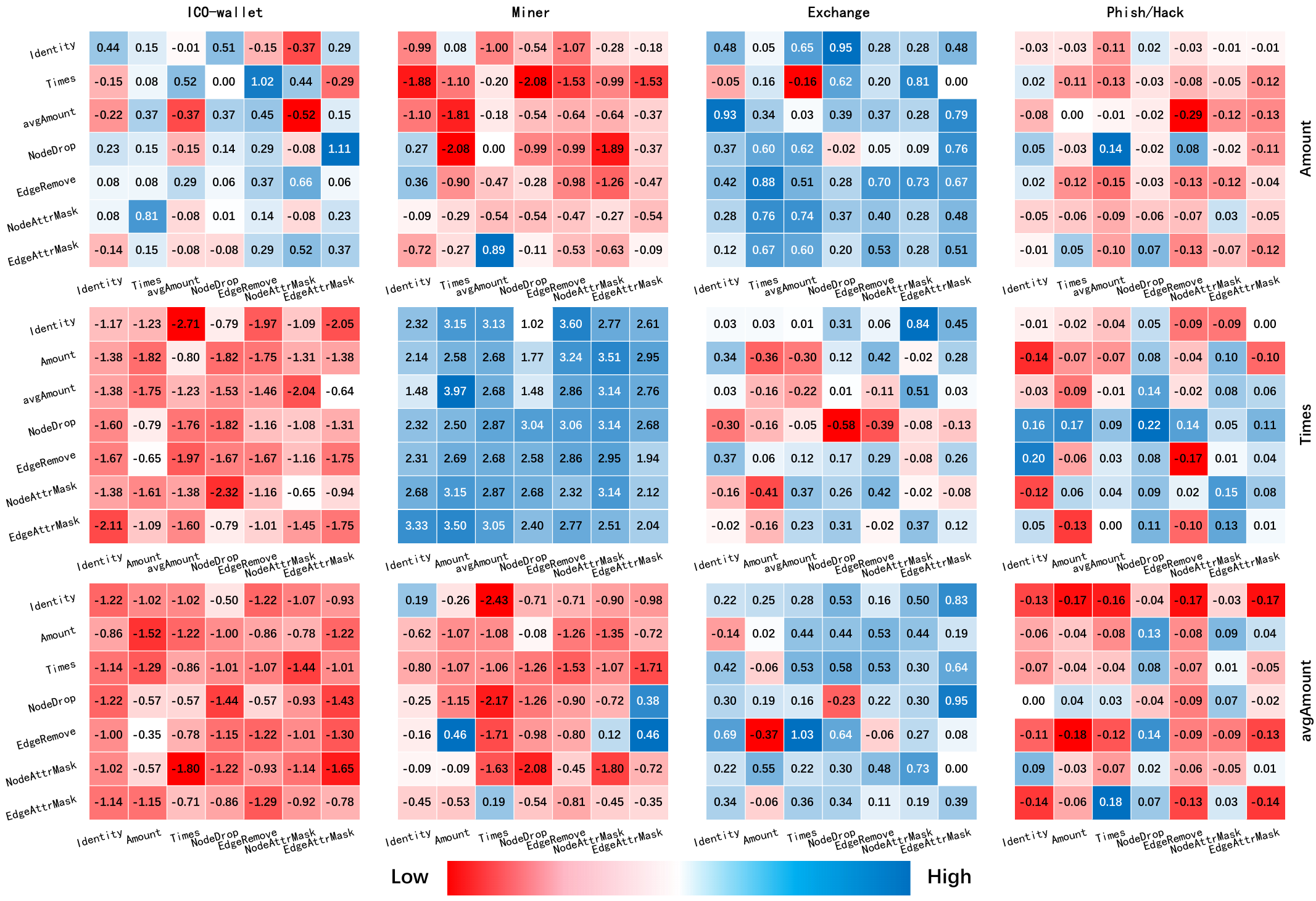}
  \caption{Account Identification F1-score gain (\%) when contrasting different augmentation pairs, compared to Ethident (w/o GC) which stands for a no-augmentation version of our framework, under all datasets. ``Identity'' represents the original view.}
  \label{fig: hotmap-aug}
\end{figure*}

\subsection{Effect of Subgraph Contrastive Learning}
We further investigate the effectiveness of subgraph contrast in our \emph{Ethident}, and list several \textbf{Obs}ervations as well as explainable analysis.

\subsubsection{\textbf{Obs. 3. Graph augmentation is crucial, and structure-level augmentation seems to benefit more}}
We first apply various pairs of augmentation views to all datasets, as illustrated in Fig.~\ref{fig: hotmap-aug}, and obtain the performance gain of \emph{Ethident} compared with \emph{Ethident (w/o GC)} which stands for a no-augmentation version of our framework (i.e., identifying accounts by using our encoder and a followed prediction head). 
Overall, it seems more likely to yield positive gain by using either ``NodeDrop'' or ``EdgeRemove'' as one of the augmented views, when compared to other augmentation pairs.
In addition, for exchange accounts that frequently call various contracts, there will be more non-zero values in their node features, making attribute masking an effective augmentation strategy as well.
Finally, we note that the combination of various augmentation views is sensitive to subgraph datasets with different sampling strategies, i.e., the performance gain of our \emph{Ethident} with the same augmentation pairs varies largely across datasets with different sampling strategies, which encourages adaptive selections of sampling strategies and augmentation combinations in future work.

\begin{figure}[htp]
	\centering
  \includegraphics[width=\linewidth]{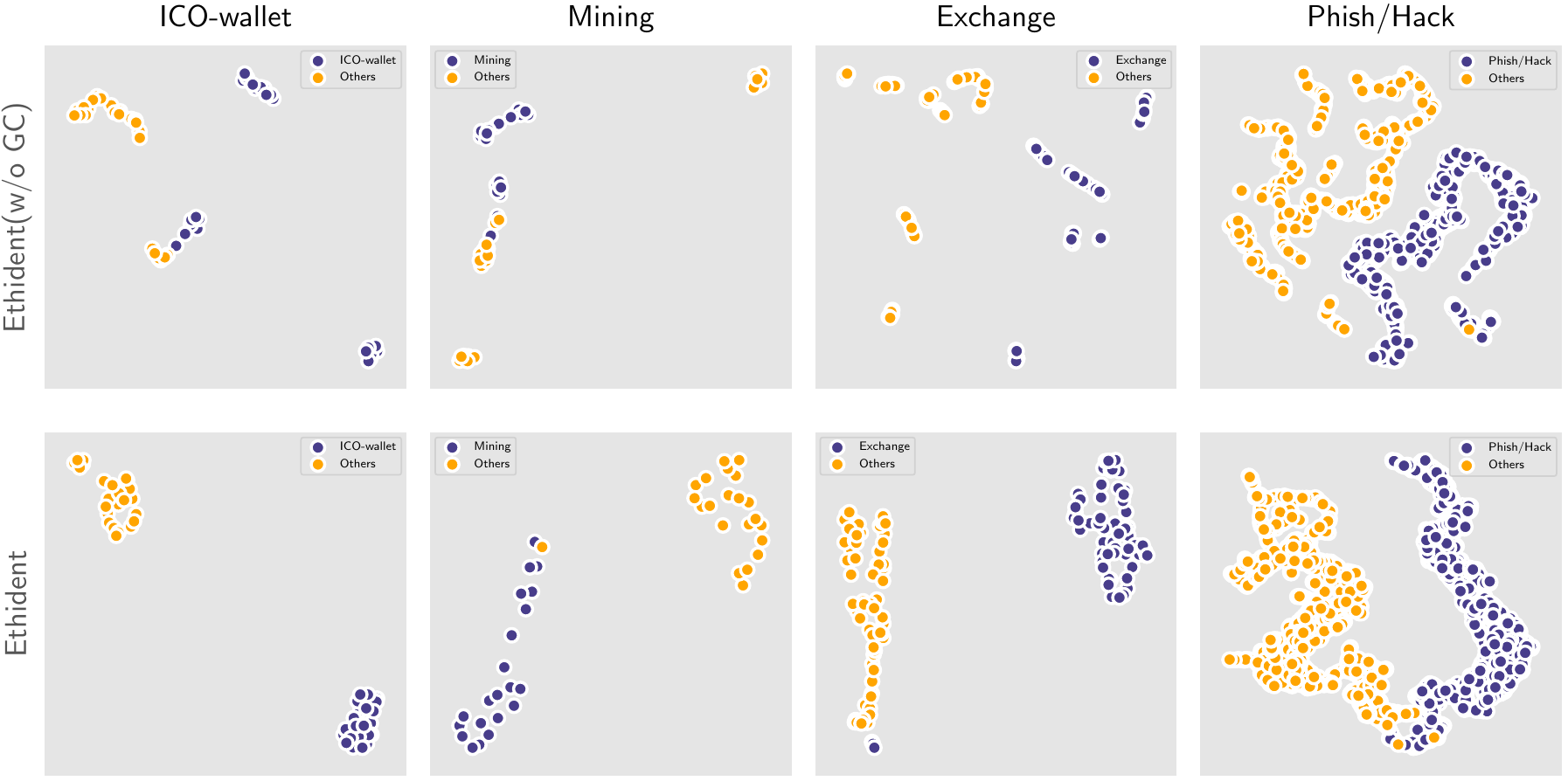}
  \caption{The UMAP visualization of the subgraph embeddings learned by Ethident with and without subgraph contract.}
  \label{fig: umap}
\end{figure}

\begin{table}[htp]
	\renewcommand\arraystretch{1.1}
	\centering
	\caption{Comparison of encoder performance with and without subgraph-level attentive pooling.}
	\label{tb: no-attpooling}
	\resizebox{\linewidth}{!}{%
	\begin{threeparttable}
	\begin{tabular}{clccc} 
	\hline\hline
	\multirow{2}{*}{Dataset}       & \multicolumn{1}{c}{\multirow{2}{*}{Method}}            & \multicolumn{3}{c}{Sampling Strategy}                                                                               \\ 
	\cline{3-5}
								   & \multicolumn{1}{c}{}                                   & Amount                                  & Times                                   & averAmount                            \\ 
	\hline
	\multirow{3}{*}{Eth-ICO}       & \emph{HGATE} (\footnotesize{w/o AttPooling})\tnote{1}  & 92.81\footnotesize{$\pm$0.034}          & 93.30\footnotesize{$\pm$0.039}          & 93.57\footnotesize{$\pm$0.034}           \\
								   & \emph{HGATE}                                           & \textbf{93.02\footnotesize{$\pm$0.029}} & \textbf{93.36\footnotesize{$\pm$0.032}} & \textbf{94.38\footnotesize{$\pm$0.028}}  \\
								   & gain                                                   & 0.23\%                                  & 0.06\%                                  & 0.87\%                                \\ 
	\hline
	\multirow{3}{*}{Eth-Mining}     & \emph{HGATE} (\footnotesize{w/o AttPooling})           & \textbf{86.21\footnotesize{$\pm$0.055}} & \textbf{84.38\footnotesize{$\pm$0.072}} & \textbf{85.38\footnotesize{$\pm$0.030}}  \\
								   & \emph{HGATE}                                           & 85.62\footnotesize{$\pm$0.060}          & 83.68\footnotesize{$\pm$0.080}          & 84.91\footnotesize{$\pm$0.051}           \\
								   & gain                                                   & -0.68\%                                 & -0.83\%                                 & -0.55\%                               \\ 
	\hline
	\multirow{3}{*}{Eth-Exchange}  & \emph{HGATE} (\footnotesize{w/o AttPooling})           & 91.27\footnotesize{$\pm$0.026}          & 91.12\footnotesize{$\pm$0.027}          & 90.98\footnotesize{$\pm$0.023}           \\
								   & \emph{HGATE}                                           & \textbf{92.28\footnotesize{$\pm$0.027}} & \textbf{92.77\footnotesize{$\pm$0.027}} & \textbf{92.39\footnotesize{$\pm$0.021}}  \\
								   & gain                                                   & 1.11\%                                  & 1.81\%                                  & 1.55\%                                \\ 
	\hline
	\multirow{3}{*}{Eth-PhishHack} & \emph{HGATE} (\footnotesize{w/o AttPooling})           & 96.81\footnotesize{$\pm$0.006}          & 96.15\footnotesize{$\pm$0.009}          & 96.72\footnotesize{$\pm$0.005}           \\
								   & \emph{HGATE}                                           & \textbf{97.79\footnotesize{$\pm$0.003}} & \textbf{97.37\footnotesize{$\pm$0.004}} & \textbf{97.80\footnotesize{$\pm$0.003}}  \\
								   & gain                                                   & 1.01\%                                  & 1.27\%                                  & 1.12\%                                \\
	\hline\hline
	\end{tabular}
	\begin{tablenotes}
		\footnotesize
		\item[1] AttPooling: subgraph-level attentive pooling operation.
	  \end{tablenotes}
	  \end{threeparttable}}
\end{table}

\begin{table}[htp]
	\renewcommand\arraystretch{1.1}
	\centering
	\caption{Results of account identification on EOSIO.}
	\label{tb: eos}
	\resizebox{\linewidth}{!}{%
	\begin{tabular}{cccc} 
	\hline\hline
	\multirow{2}{*}{Method}        & \multicolumn{3}{c}{EOSIO}                                                \\ 
	\cline{2-4}       
							       & $h=1$, $K=10$                      & $h=1$, $K=20$                     & $h=2$, $K=10$                        \\ 
	\hline       
	GCN                            & 99.59$\pm$0.003                    & 99.47$\pm$0.003                   & 99.30$\pm$0.004                     \\
	GAT                            & 99.52$\pm$0.002                    & 99.68$\pm$0.002                   & 99.12$\pm$0.004                     \\
	GIN                            & 99.52$\pm$0.002                    & 99.55$\pm$0.001                   & 99.40$\pm$0.002                     \\
	$\text{I}^2\text{BGNN}$-A      & 99.47$\pm$0.002                    & 99.53$\pm$0.002                   & 99.12$\pm$0.005                     \\
	$\text{I}^2\text{BGNN}$-T      & 99.25$\pm$0.002                    & 99.62$\pm$0.002                   & 99.17$\pm$0.005                     \\
	\textbf{Ethident} (w/o GC)     & 99.58$\pm$0.003                    & 99.70$\pm$0.002                   & 99.20$\pm$0.005                     \\
	\textbf{Ethident}              & \textbf{99.75}$\pm$\textbf{0.001}  & \textbf{99.75}$\pm$\textbf{0.001} & \textbf{99.47}$\pm$\textbf{0.002}  \\
	\hline\hline
	\end{tabular}}
\end{table}

\begin{table}[htp]
	\renewcommand\arraystretch{1.1}
	\centering
	\caption{Results on the original and malfunctioning exchange \& phishing datasets.}
	\label{tb: adv}
	\resizebox{1\linewidth}{!}{%
	\begin{tabular}{c|c|c|ccc} 
	\hline\hline
	\multirow{5}{*}{\begin{tabular}[c]{@{}c@{}}Sample\\Strategy\end{tabular}} & \multirow{5}{*}{\begin{tabular}[c]{@{}c@{}}Test with\\$D_\textit{test}^\textit{ori}$ or $D_\textit{test}^\textit{mal}$\end{tabular}} & \multicolumn{4}{c}{Method}                                                                                                                                                                                                                                                      \\ 
	\cline{3-6}
									&                  & \multirow{4}{*}{\begin{tabular}[c]{@{}c@{}}Ethident\\(w/o GC)\end{tabular}} & \multicolumn{3}{c}{Ethident}                                                                                                                                                                     \\ 
	\cline{4-6}
									&                  &                                                                             & \begin{tabular}[c]{@{}c@{}}edgeRemove\\\&\\identity\end{tabular} & \begin{tabular}[c]{@{}c@{}}edgeRemove\\\&\\edgeRemove\end{tabular} & \begin{tabular}[c]{@{}c@{}}edgeRemove\\\&\\nodeDrop\end{tabular}  \\ 
	\hline
	\multirow{3}{*}{averAmount}     & Original         & 89.74$\pm$0.026     & 90.57$\pm$0.031   & 91.22$\pm$0.029    & 91.74$\pm$0.025                                                    \\
									& Malfunctioning   & 89.56$\pm$0.038     & 91.32$\pm$0.029   & 91.24$\pm$0.031    & 91.22$\pm$0.030                                                    \\
									& Loss             & -0.20\%             & 0.83\%            & 0.02\%             & -0.57\%                                                    \\ 
	\hline
	\multirow{3}{*}{Amount}         & Original         & 89.09$\pm$0.035     & 90.70$\pm$0.033   & 89.39$\pm$0.039    & 90.86$\pm$0.024                                                    \\
									& Malfunctioning   & 89.28$\pm$0.034     & 90.13$\pm$0.027   & 89.92$\pm$0.030    & 90.96$\pm$0.030                                                    \\
									& Loss             & 0.21\%              & -0.63\%           & 0.59\%             & 0.11\%                                                         \\ 
	\hline
	\multirow{3}{*}{Times}          & Original         & 86.97$\pm$0.029     & 87.92$\pm$0.029   & 87.51$\pm$0.027    & 88.83$\pm$0.023                                                    \\
									& Malfunctioning   & 87.21$\pm$0.034     & 87.40$\pm$0.035   & 87.48$\pm$0.028    & 88.33$\pm$0.031                                                    \\
									& Loss             & 0.28\%              & -0.59\%           & -0.03\%            & -0.56\%                                                        \\
	\hline\hline
	\end{tabular}}
\end{table}

\subsubsection{\textbf{Obs. 4. Contrastive self-supervision improves the generalization of model in account feature learning}}
We utilize the UMAP~\cite{mcinnes2018umap-software} to visualize the subgraph embeddings learnt by \emph{Ethident} and \emph{Ethident (w/o GC)} in Fig.~\ref{fig: umap}, where different colors mean different labels. 
Compared with \emph{Ethident (w/o GC)} which only uses prediction loss $\mathcal{L}_\text{pred}$, more obvious inter-class separability and intra-class compactness are achieved after applying the contrastive constraint $\mathcal{L}_\text{self}$, which illustrates its effectiveness on learning the behavior pattern differences.
Moreover, \emph{Ethident} separates different patterns with relatively clearer boundaries, suggesting that contrastive self-supervision can effectively improve the generalization of models when training with scarce labels.

\subsection{More Analysis}
\subsubsection{Impact of Subgraph-level Attentive Pooling}
To illustrate the effectiveness of subgraph-level attentive pooling, we compare the performance of our encoder with and without this module, as reported in Table~\ref{tb: no-attpooling}.
We observe that the encoder with \emph{AttPooling} achieves better performance on 3 out of 4 categories of accounts, validating the effectiveness of our proposal.
For the exception that \emph{AttPooling} brings a negative gain on \emph{Mining} subgraphs, we speculate that a mining pool organization usually behaves very differently from an individual miner whose transaction behavior has no significant relationship with the mining pool, so aggregating information from neighbors may interfere with the characterization of mining pools' behavior patterns.

\begin{figure*}[htp]
	\centering
  \includegraphics[width=\textwidth]{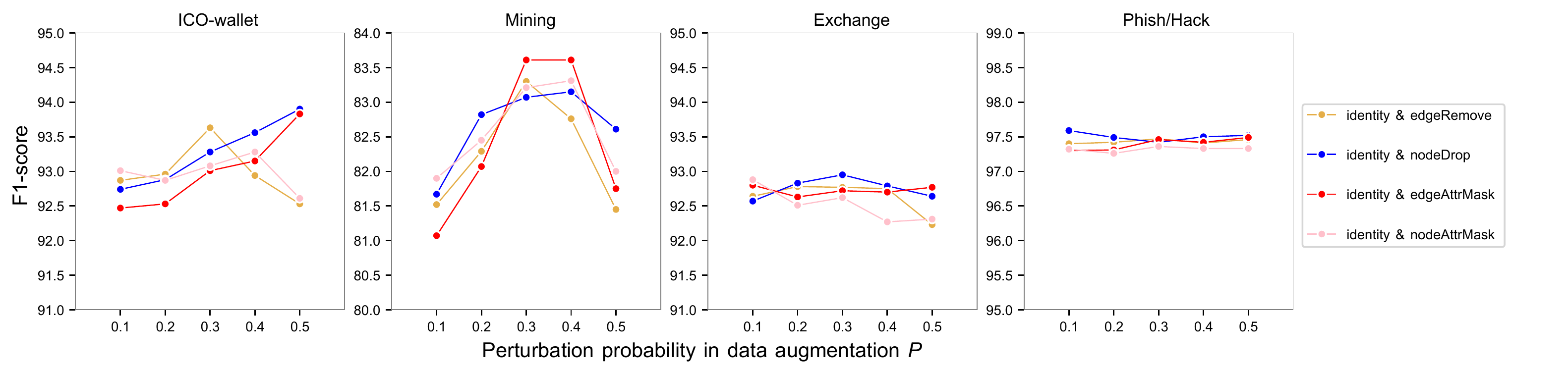}
  \caption{Impact of perturbation probability in data augmentation ($\mathcal{P}$).}
  \label{fig: line-DA-p}
\end{figure*}

\begin{figure}[htp]
	\centering
  \includegraphics[width=\linewidth]{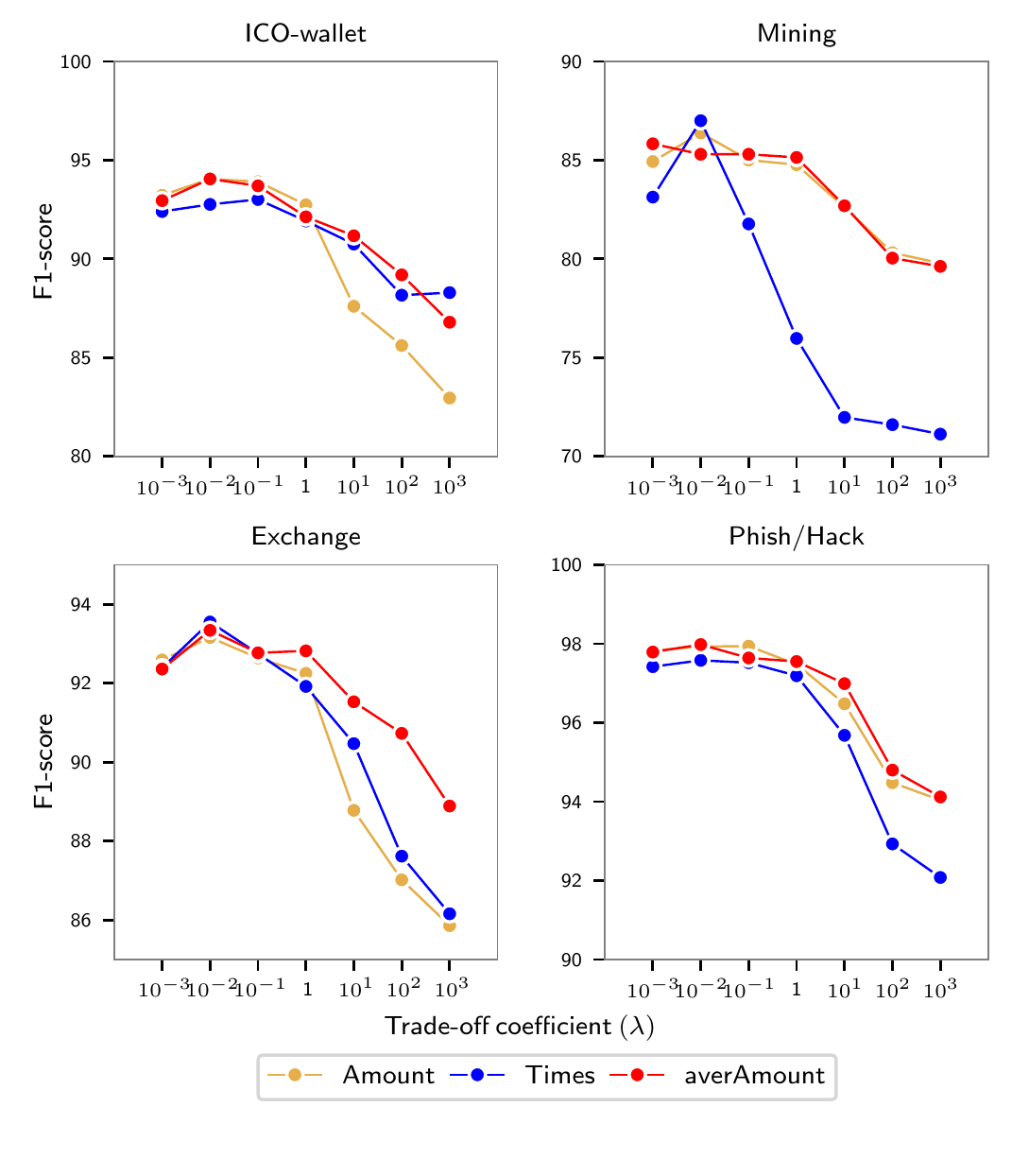}
  \caption{Impact of tradeoff coefficient ($\lambda$).}
  \label{fig: line-lambda}
\end{figure}

\subsubsection{Impact of Perturbation Probability}
We continue to analyze the impact of perturbation probability $\mathcal{P}$ in data augmentation.
We use a view for data augmentation (in the form of ``Identity \& DA'') and vary $\mathcal{P}$ in $\{0.1, 0.2, \cdots, 0.5\}$, the results are shown in Fig.~\ref{fig: line-DA-p}.
Combined with the statistics in Table~\ref{tb: subgraph}, we have drawn the following conclusions:
(1) Datasets with larger amounts of samples are more robust to variation in perturbation probability;
(2) Datasets with larger average sample scales (in terms of \textit{Avg.} $|N|$ and \textit{Avg.} $|V|$) are less sensitive to variation in perturbation probability;
(3) An effective and reasonable selection interval for the perturbation probability could be [0.1, 0.3].
Finally, we observe that our method still performs well when the perturbation is large, which is likely to benefit from the attentive graph pooling.

\subsubsection{Impact of Loss Tradeoff}
Here we analyze the impact of the trade-off coefficient $\lambda$ which controls the contribution of subgraph contrast.
As we can see from Fig.~\ref{fig: line-lambda}, our \emph{Ethident} achieves relatively better performance when $\lambda$ is less than 1, which meets our intuition.
We treat subgraph contrast as a pretext task or a regularization to subgraph classification.
When the coefficient of regularization is greater than 1, the classification task cannot be fully optimized, failing to learn the task-related features.

\subsubsection{Tradeoffs between Performance and Consumption}
Since the subgraph extraction allows for mini-batch training of our framework, greatly reducing computational consumption and time cost. 
Here we further investigate the tradeoffs between performance and consumption under different sampling scales, as shown in Fig.~\ref{fig: line-size}.
Since the performance of 2-hop subgraphs significantly outperforms that of 1-hop subgraphs, we just draw the consumption curves of 2-hop subgraph.  
We can first observe that the memory and time consumption increase almost linearly with the sample scale, while the performance converges when the sample scale increases to a certain extent.
We then use the ``$\frac{\text{performance}}{\text{consumption}}$'' metric to roughly analyze the tradeoff between them.
Note that a larger ``$\frac{\text{performance}}{\text{consumption}}$'' metric generally indicates better performance with less consumption.
After observation and calculation, we finally conclude that an appropriate parameter setting of sample scale could be $h=2$ and $K \in [8,10]$.

\subsubsection{Generalization Application to Other Cryptocurrency}
To verify the generalization of our framework on other cryptocurrencies, we collect transaction data of another on-chain cryptocurrency EOSIO and deploy related experiments.
We first construct an account interaction graph including 944,865 nodes and 10,435,037 edges, in which the node features ($\mathbf{X} \in \mathbb{R}^{n\times 1216}$) are constructed by the contract calling information and account name restriction mechanism, and the edge features are constructed in the same way as in Sec.~\ref{sec: edge-feature}. 
We then collect 2000 target accounts, half of which are normal accounts and half are bot accounts, and our goal is to determine the identity of these accounts, that is, normal accounts or bot accounts.
We apply our \emph{Ethident} framework to achieve the account identification on the EOSIO dataset, and the experimental settings are similar to that in Sec.~\ref{sec: exp}.
Table~\ref{tb: eos} report the account identification results on EOSIO dataset.
As we can see, our \emph{Ethident} still achieves the state-of-the-art identification performance on the EOSIO dataset, showing a good generalization to other cryptocurrencies.

\subsubsection{Generalization Evaluation on Malfunctioning Exchange Accounts}
Accounts in the same broad category share common transaction patterns, but also have their own particularities, so that they can generally still be classified at a more fine-grained level. 
For example, the malfunctioning exchange accounts which have different transaction patterns from normal ones are common in real Ethereum, and still belong to `Exchange' accounts. However, these malfunctioning exchange accounts which only receive amount from trader accounts but do not send amount back to traders may be more likely to be misclassified as a `Phishing' account rather than an `Exchange' account, as it has a more similar transaction pattern to the former.
Here, we conduct experiments to validate whether our \emph{Ethident} model can successfully identify the malfunctioning exchange account as an `Exchange' account instead of a `Phishing' account.
\begin{itemize}
	\item Randomly select the same number of `Exchange' accounts and `Phishing' accounts, and extract their transaction subgraphs, yielding a new dataset of 386 account subgraphs. Split the new dataset into training $D_\textit{train}$, validation $D_\textit{val}$ and testing $D_\textit{test}^\textit{ori}$ sets with a proportion of $1:1:1$.
	\item Generate the malfunctioning exchange account by removing the target exchange account's outgoing edges from account subgraph. In this way, we yield the malfunctioning testing set $D_\textit{test}^\textit{mal}$ containing phishing accounts and malfunctioning exchange accounts.
	\item Train \emph{Ethident} model to determine whether an account is an `Exchange' account or a `Phishing' account using $D_\textit{train}$ and $D_\textit{val}$.
	\item Evaluate the performance of the model in determining whether an account is an `Exchange' account or a `Phishing' account using $D_\textit{test}^\textit{ori}$ and $D_\textit{test}^\textit{mal}$ respectively.
\end{itemize}
As we can see from Table~\ref{tb: adv}, our \emph{Ethident} models still achieve powerful performance in identifying indistinguishable `Phishing' accounts and `Malfunctioning Exchange' accounts. 
Compared with the original results in $D_\textit{test}^\textit{ori}$, our \emph{Ethident} show -0.63\% $\sim$ 0.83\% performance fluctuations in malfunctioning testing set $D_\textit{test}^\textit{mal}$, which is a rational and normal performance jitter. This phenomenon suggests that our model has almost no performance loss in distinguishing between phishing accounts and malfunctioning exchange accounts, showing strong robustness and generalization.

\section{Conclusion}
Financial security has become a top priority in the blockchain ecosystem.
This paper provides a new perspective on account de-anonymization, and proposes a behavior-aware Ethereum account identification framework that integrates hierarchical graph attention and self-supervision mechanism, to effectively characterize the behavior patterns of different accounts.
Extensive experiments on Ethereum datasets demonstrate the superiority of our framework in terms of state-of-the-art performance and powerful generalization.
Furthermore, our framework also has a good transferability to other blockchain platforms like Bitcoin and EOSIO, which will be discussed in future work.

\begin{table*}[htp]
	\centering
	\renewcommand\arraystretch{1.2}
	\caption{Statistics of the average of manual feature for various accounts in Ethereum.}
	\label{tb: manual}
	\resizebox{\textwidth}{!}{%
	\begin{tabular}{lcccccr} 
		\hline\hline
		Manual Features                     & Phish-Hack & Exchange   & Mining    & ICO-Wallets & Common      & Definition  \\ 
		\hline        
		\emph{active\_days}                 & 76.94      & 703.35     & 595.98    & 547.84      & 14.49       & the number of active days of the account.  \\
		\emph{total\_received}              & 110.84     & 1551629.77 & 5470.67   & 6642.11     & 245.38      & the total amount of Ether received by the account.  \\
		\emph{num\_received\_tx}            & 27         & 88490.39   & 68.55     & 279.86      & 4.13        & the number of transactions with Ether received by the account. \\
		\emph{inter\_acct\_received}        & 23.29      & 29985.26   & 13.6      & 218.77      & 0.4         & the number of accounts sending Ether to the target account.  \\
		\emph{total\_output}                & 124.03     & 2309107.24 & 367339.66 & 33824.02    & 370.21      & the amount of Ether spent by the account.  \\
		\emph{num\_output\_tx}              & 29.91      & 88285.11   & 818877.92 & 62.84       & 4.56        & the number of transactions that the account has spent Ether.  \\
		\emph{inter\_acct\_output}          & 8.18       & 46692.69   & 16825.85  & 37.53       & 1.48        & the number of accounts receive Ether from the target account.  \\
		\emph{avg\_received}                & 29.52      & 2002.55    & 49.79     & 854.35      & 7.06        & the average amount of Ether received by the account.  \\
		\emph{avg\_received\_day}           & 11.66      & 1748.8     & 6.55      & 11.36       & 4.93        & the average amount of Ether received by the account per day.  \\
		\emph{avg\_received\_tx\_day}       & 1.51       & 113.5      & 0.16      & 0.49        & 0.03        & the number of transactions with Ether received by the account per day.  \\
		\emph{avg\_output}                  & 30.58      & 1453.75    & 249.95    & 4399.06     & 9.46        & the average amount of Ether spent by the account. \\
		\emph{avg\_output\_day}             & 14.66      & 2213.77    & 389.08    & 80.3        & 5.39        & the average amount of Ether spent by the account per day.  \\
		\emph{avg\_output\_tx\_day}         & 0.66       & 101.26     & 633.21    & 0.2         & 0.03        & the average number of transactions with Ether spent by the account per day.  \\
		\emph{times\_contract\_called}      & 11.68      & 66682.01   & 61002.03  & 690.93      & 1.52        & the number of times the account calls the smart contract.  \\
		\emph{times\_contract\_called\_day} & 0.5        & 82.66      & 41.84     & 1.07        & 0.02        & the number of times the account calls the smart contract per day.  \\
		\emph{num\_contract\_called}        & 3.29       & 2031.6     & 1256.31   & 3.26        & 0.13        & the number of contracts called by the account.  \\
		\hline\hline
	\end{tabular}
	}
\end{table*}

\section*{Acknowledgments}
This work was partially supported by the Key R\&D Program of Zhejiang under Grant 2022C01018, by the National Key R\&D Program of China under Grant 2020YFB1006104, by the National Natural Science Foundation of China under Grant 61973273, by the Zhejiang Provincial Natural Science Foundation of China under Grant LR19F030001, and by the Major Key Project of PCL under Grants PCL2022A03, PCL2021A02 and PCL2021A09.

{\appendices

\section{Manual Feature Details}\label{app: manual}
Manual feature engineering is the most common and simplest way for account identification. 
According to the characteristics of raw Ethereum data and prior knowledge, we design 16 manual features for Ethereum accounts, as shown in Table~\ref{tb: manual}.

}

\bibliographystyle{IEEEtran}
\bibliography{ref}


\vspace{-35pt}


\begin{IEEEbiography}[{\includegraphics[width=1in,height=1.25in,clip,keepaspectratio]{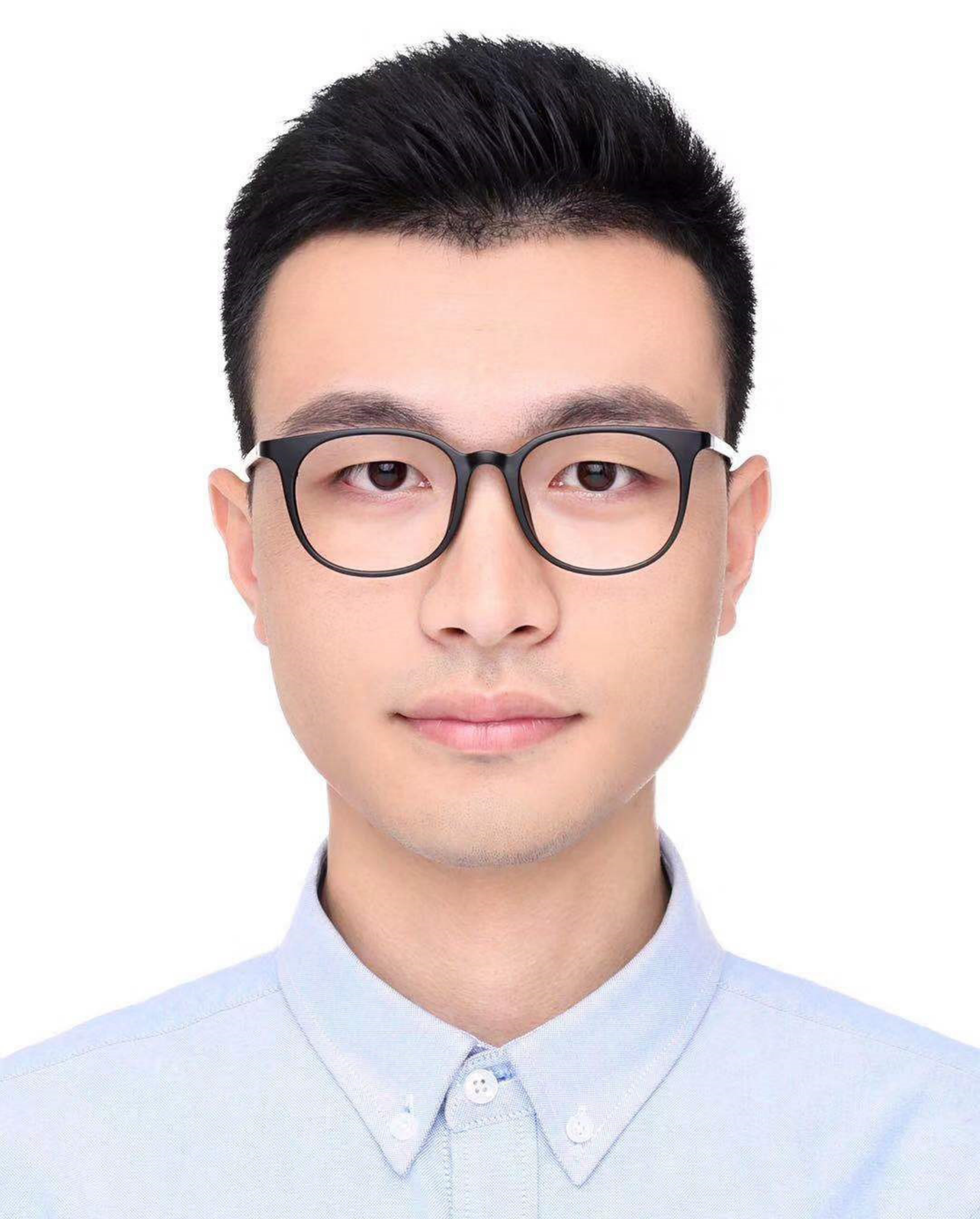}}]{Jiajun Zhou}
	received the BS degree in automation from the Zhejiang University of Technology, Hangzhou, China, in 2018, where he is currently pursuing the Ph.D degree in control theory and engineering with the College of Information and Engineering.
	His current research interests include graph data mining and deep learning, especially for graph self-supervised learning and blockchain data analytics.
\end{IEEEbiography}
\vspace{-35pt}
\begin{IEEEbiography}[{\includegraphics[width=1in,height=1.25in,clip,keepaspectratio]{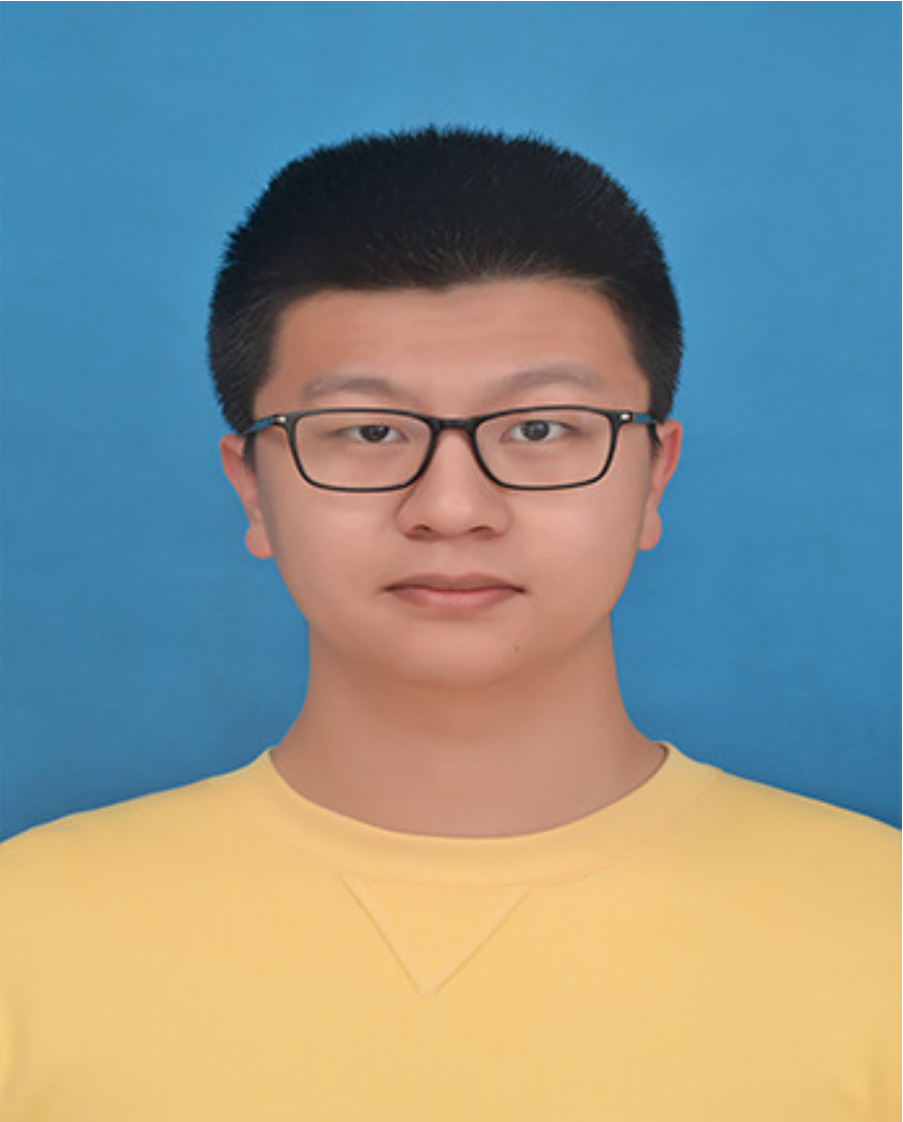}}]{Chenkai Hu}
	is currently pursuing the bachelor’s degree in automation at Zhejiang University of Technology, HangZhou, China. His current research interests include data mining in blockchain.
\end{IEEEbiography}
\vspace{-35pt}
\begin{IEEEbiography}[{\includegraphics[width=1in,height=1.25in,clip,keepaspectratio]{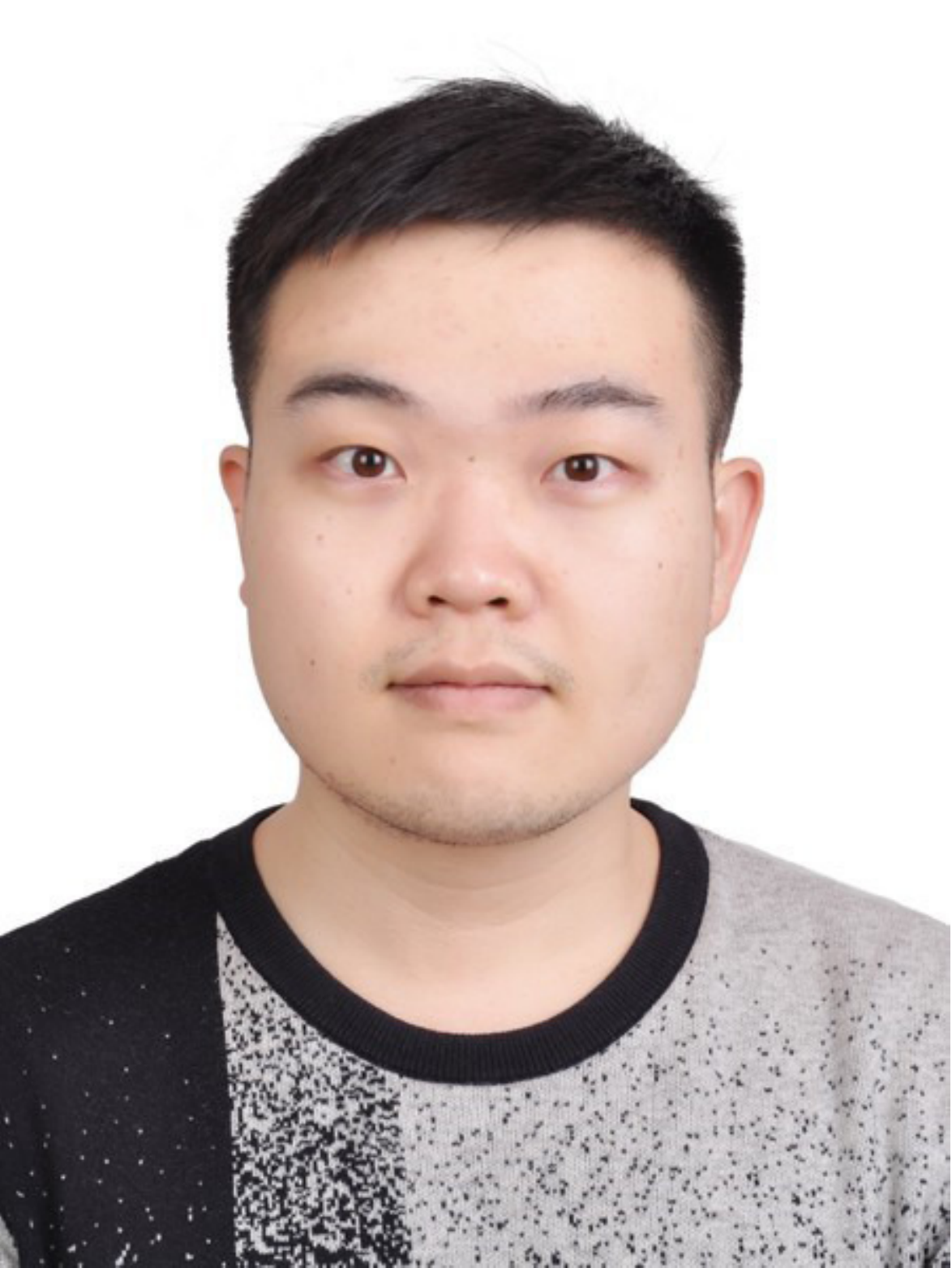}}]{Jianlei Chi}
	received the B.S. degree in computer science and technology from Harbin Engineering University, China, 2014, and the Ph.D. degree in computer science and technology in 2022 from Xi’an Jiaotong University, China. He is currently an assistant professor at Hangzhou Research Institute of Xidian University. His research interests include trustworthy software, software engineering, program analysis and machine learning.
\end{IEEEbiography}
\vspace{-35pt}
\begin{IEEEbiography}[{\includegraphics[width=1in,height=1.25in,clip,keepaspectratio]{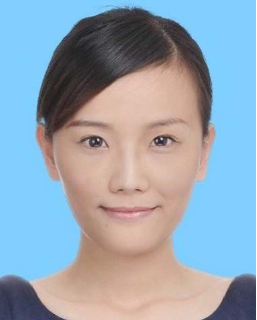}}]{Jiajing Wu}
	(Senior Member, IEEE) received the Ph.D. degree from The Hong Kong Polytechnic University, Hong Kong, in 2014. In 2015, she joined Sun Yat-sen University, Guangzhou, China, where she is currently an Associate Professor. Her research interests include blockchain, graph mining, and network science.Dr. Wu was awarded the Hong Kong Ph.D. Fellowship Scheme during her Ph.D. degree in Hong Kong from 2010 to 2014. She also serves as an Associate Editor for IEEE TRANSACTIONS ON CIRCUITS AND SYSTEMS II: EXPRESS BRIEFS.
\end{IEEEbiography}
\vspace{-35pt}
\begin{IEEEbiography}[{\includegraphics[width=1in,height=1.25in,clip,keepaspectratio]{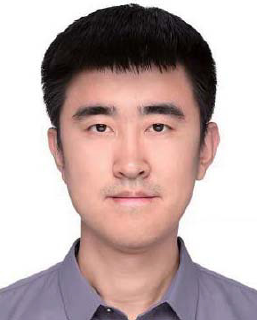}}]{Meng Shen}
	(Member, IEEE) received the B.Eng. degree in computer science from Shandong University, Jinan, China, in 2009, and the Ph.D. degree in computer science from Tsinghua University, Beijing, China, in 2014. He is currently an Associate Professor with Beijing Institute of Technology, Beijing. He has authored over 50 papers in top-level journals and conferences, such as ACM SIGCOMM, IEEE JOURNAL ON SELECTED AREAS IN COMMUNICATIONS (JSAC), and IEEE TRANSACTIONS ON INFORMATION FORENSICS AND SECURITY (TIFS). His research interests include data privacy and security, blockchain applications, and encrypted traffic classification. He received the Best Paper Award from IEEE/ACM IWQoS 2021. He was selected by the Beijing Nova Program 2020 and the winner of the ACM SIGCOMM China Rising Star Award in 2019. He has guest edited Special Issues on Emerging Technologies for Data Security and Privacy in IEEE Network and IEEE INTERNET OF THINGS JOURNAL.
\end{IEEEbiography}
\vspace{-35pt}
\begin{IEEEbiography}[{\includegraphics[width=1in,height=1.25in,clip,keepaspectratio]{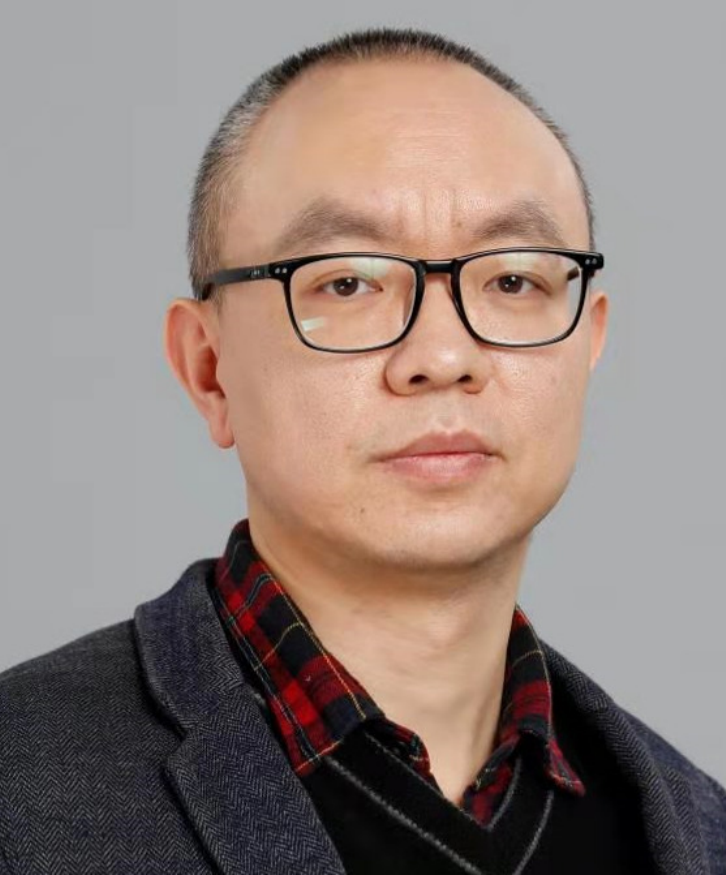}}]{Qi Xuan}(M'18) received the BS and PhD degrees in control theory and engineering from Zhejiang University, Hangzhou, China, in 2003 and 2008, respectively. He was a Post-Doctoral Researcher with the Department of Information Science and Electronic Engineering, Zhejiang University, from 2008 to 2010, respectively, and a Research Assistant with the Department of Electronic Engineering, City University of Hong Kong, Hong Kong, in 2010 and 2017. From 2012 to 2014, he was a Post-Doctoral Fellow with the Department of Computer Science, University of California at Davis, CA, USA. He is a senior member of the IEEE and is currently a Professor with the Institute of Cyberspace Security, College of Information Engineering, Zhejiang University of Technology, Hangzhou, China. His current research interests include network science, graph data mining, cyberspace security, machine learning, and computer vision.
\end{IEEEbiography}


\end{document}